\documentclass[12pt,preprint]{aastex}
\usepackage{times}
\usepackage{graphicx}
\usepackage{epstopdf}
\usepackage{natbib}
\graphicspath{/Users/garymelnick/Desktop/astro-ph2}

\newcommand{\ltsima}{\stackrel{\textstyle <}{\sim}}
\newcommand{\simlt}{\scriptsize{\raisebox{-2pt}{$\ltsima$}}\normalsize}
\newcommand{\gtsima}{\stackrel{\textstyle >}{\sim}}
\newcommand{\simgt}{\scriptsize{\raisebox{-2pt}{$\gtsima$}}\normalsize}

\renewcommand{\thepage}{\rm\arabic{page}}

\newcommand{\asec}{$^{\prime\prime}$}
\newcommand{\amin}{$^{\prime}$}
\newcommand{\mh}{H$_2$}
\newcommand{\nh}{n{(H$_2$)}}

\newcommand{\water}{H$_2$O}
\newcommand{\nwater}{H$_2^{\; 16}$O}

\newcommand{\rv}{$R_{\rm V}$}
\newcommand{\um}{$\:\mu$m}
\newcommand{\kms}{~km~s$^{{-1}}$}
\newcommand{\cmc}{cm$^{-3}$}

\newcommand{\etal}{{\em et$\;$al.}}

\makeatletter
\def\fnum@figure{{Fig.~a}}
\long\def\@makecaption#1#2{
 \vskip 10pt 
 \setbox\@tempboxa\hbox{#1. #2}
 \ifdim \wd\@tempboxa >\hsize \unhbox\@tempboxa\par \else \hbox
to\hsize{\hfil\box\@tempboxa\hfil} 
 \fi}

\makeatother

\makeatletter
\def\ps@myheadings{\let\@mkboth\@gobbletwo
\def\@oddhead{\hbox{}\sl\rightmark \hfil \rm \thepage}%
\def\@oddfoot{}\def\@evenhead{\thepage\hfil\sl\leftmark\hbox {}}%
\def\@evenfoot{}\def\sectionmark##1{}\def\subsectionmark##1{}}
\makeatother

\makeatletter
\def\fnum@figure{{Fig.~\thefigure}}
\long\def\@makecaption#1#2{
 \vskip 10pt 
 \setbox\@tempboxa\hbox{#1. #2}
 \ifdim \wd\@tempboxa >\hsize \unhbox\@tempboxa\par \else \hbox
to\hsize{\hfil\box\@tempboxa\hfil} 
 \fi}

\makeatother

\shorttitle{Extended Hot Water in NGC 2071}
\shortauthors{Melnick et al.}

\renewcommand{\baselinestretch}{1.25}

\begin{document}

\title{Detection of Extended Hot Water in the Outflow from NGC$\:$2071}
\author{Gary J. Melnick, Volker Tolls} 
\affil{Harvard-Smithsonian Center for Astrophysics, 60 Garden Street, 
Cambridge, MA 02138}
\email{gmelnick@cfa.harvard.edu, vtolls@cfa.harvard.edu}
\author{David A. Neufeld, Yuan Yuan, Paule Sonnentrucker}
\affil{Department of Physics and Astronomy, Johns Hopkins University,
3400 North Charles Street, Baltimore, MD 21218}
\email{neufeld@pha.jhu.edu, yuanyuan@pha.jhu.edu, sonnentr@pha.jhu.edu}
\author{Dan M. Watson}
\affil{Department of Physics and Astronomy, University of Rochester, 
Rochester, NY 14627}
\email{dmw@pas.rochester.edu}
\author{Edwin A. Bergin}
\affil{Department of Astronomy, University of Michigan, 825 Dennison 
Building, Ann Arbor, MI 48109}
\email{ebergin@umich.edu}
\and
\author{Michael J. Kaufman}
\affil{Department of Physics and Astronomy, San Jose State University, 
One Washington Square, San Jose, CA 95192-0106}
\email{mkaufman@email.sjsu.edu}
\author{}
\author{}
\author{}
\author{}
\author{}
\author{}

\begin{abstract}
We report the results of spectroscopic mapping observations carried out toward 
a $\sim\,$1\amin$\times$1\amin\ region within the northern lobe of the
outflow from NGC$\:$2071 using the Infrared Spectrograph (IRS) of the 
{\em Spitzer Space Telescope}.  These observations covered the 5.2$-$37\um\ spectral region
and have led to the detection of a number of ionic, atomic, and molecular lines,
including fine-structure emission of
Si$^+$, Fe$^+$, S$^{++}$, S, the S(0)-S(7) pure rotational lines of 
\mh, the R(3) and R(4) transitions of HD, and at least 11 transitions of \water.  In
addition, the 6.2, 7.4, 7.6, 7.9, 8.6 and 11.3\um\ PAH emission bands were also 
observed and several transitions of OH were tentatively detected.
Most of the detected line transitions were strong enough to map including, for the
first time, three transitions of hot \water.  
We find that: (1) the water emission is extended; (2) the extended
emission is aligned with the outflow; and, (3) the spatial distribution of the water 
emission generally follows that observed for H$_2$.  
The {\em Spitzer}/IRS data suggest that both dissociative and non-dissociative
shocks are present in the portion of the outflow observed here.  Most of H$_2$,
H$_2$O S(2)-S(7), HD, and S$\,$I emission detected likely arises from behind the 
non-dissociative shock, while the Si$\,$II, Fe$\,$II, and most of the H$_2$ S(1)
emission originate from behind the dissociative shock.  
Based on the measured line intensities, we derive an HD abundance 
relative to H$_2$ of 1.1$-$1.8$\times$10$^{-5}$
and an \water\ number density of 12$\,-\,$29~\cmc.  The \mh\ density in the
water-emitting region is not well constrained by our observations, but is likely between
3$\times$10$^4$ and 10$^6$~\cmc, yielding an \water\ abundance relative to \mh\
of between 2$\times$10$^{-5}$ and 6$\times$10$^{-4}$.  Future observations
planned for the {\em Herschel Space Observatory} should greatly improve the density
estimate, and thus our knowledge of the \water\ abundance, for the water-emitting 
regions reported here.  Finally, we note a possible departure from the \water\
ortho-to-para ratio of 3:1 expected for water formed in hot post-shocked gas,
suggesting that a significant fraction of the water vapor we detect may arise
from \water\ sputtered from cold dust grains.
\end{abstract}

\keywords{ISM: Abundances, ISM: Jets and Outflows, ISM: Molecules, Stars: Formation, Stars: Winds, Outflows}

\pagebreak

\setcounter{page}{1}
\setlength{\textheight}{8.50in}

\pagestyle{myheadings}
\markright{\hfill{\hbox{\hss\rm Page \rm }}}

\renewcommand{\baselinestretch}{1.22}

\section{\bf Introduction}

\vspace{-1mm}

Outflows from young stellar objects provide a natural laboratory toward which
it is possible to test the predictions of chemical models thought to apply to
warm ($T\,$\simgt$\,$300K), dense (\nh$\,$\simgt$\,$10$^4$~\cmc) gas.
Among these predictions is that water will be formed rapidly in the warm gas
via a set of neutral-neutral reactions \citep[cf.][]{eanddj78, eandw78}.
Moreover, these reactions can potentially convert all of the oxygen not 
locked in CO into \water\ \citep[cf.][]{bergin98}.  
Indeed, the gas-phase water abundances toward shock-heated gas 
associated with outflows is found to be higher than that within cold, quiescent 
molecular clouds \citep[cf.][]{harwit98, melnick00, neufeld00, nisini00,
benedettini02}, although the inferred abundances rarely approach 
[H$_2$O]/[H$_2$]$\,\sim\,$10$^{-4}$
as would be expected for the full conversion of all oxygen not in CO into \water.
In addition, an unknown fraction of the elevated gas-phase water abundance is 
likely due to the removal of water-ice from grains that pass through
nondissociative shocks \citep{draine83}.  
Based on a detailed analysis of ground-state ortho-water 
spectra obtained by the {\em Submillimeter Wave Astronomy Satellite} (SWAS) 
toward 18 outflow sources, \citet{franklin08} propose that the 
lower-than-expected water abundances within these sources is due to the small 
fraction of outflow material that actually passes through strong shocks.  This
also implies that the beam filling factor of regions with highly-abundant water 
is lower than previously assumed.

Unfortunately, these previous observations lacked the spatial
resolution to determine the detailed water distribution. Thus, the 
interpretation was hampered by uncertainties in both the fraction of the
beam filled with \water\ emission and the physical conditions in the 
\water-emitting gas
as might be revealed by co-spatial emission from \mh.
With beam sizes that range between 1.8\asec\ and 5.5\asec, depending on 
wavelength, and high sensitivity, the Infrared Spectrometer (IRS) aboard 
the {\em Spitzer Space Telescope} has enabled the study of
\water, \mh, and HD with unprecedented detail.  In this paper, we describe
observations of the outflow source NGC 2071 and report the
first detection of spatially extended, hot \water\ vapor with {\em Spitzer}/IRS. 
In addition to \water, we have also mapped the ground-vibrational state
distribution of \mh, as well as low-lying rotational and fine-structure
emission from HD, S$\,$I, Fe$\,$II, and Si$\,$II.

\begin{figure}[t]
\vspace{1.1in}
\centering
\includegraphics[scale=0.75]{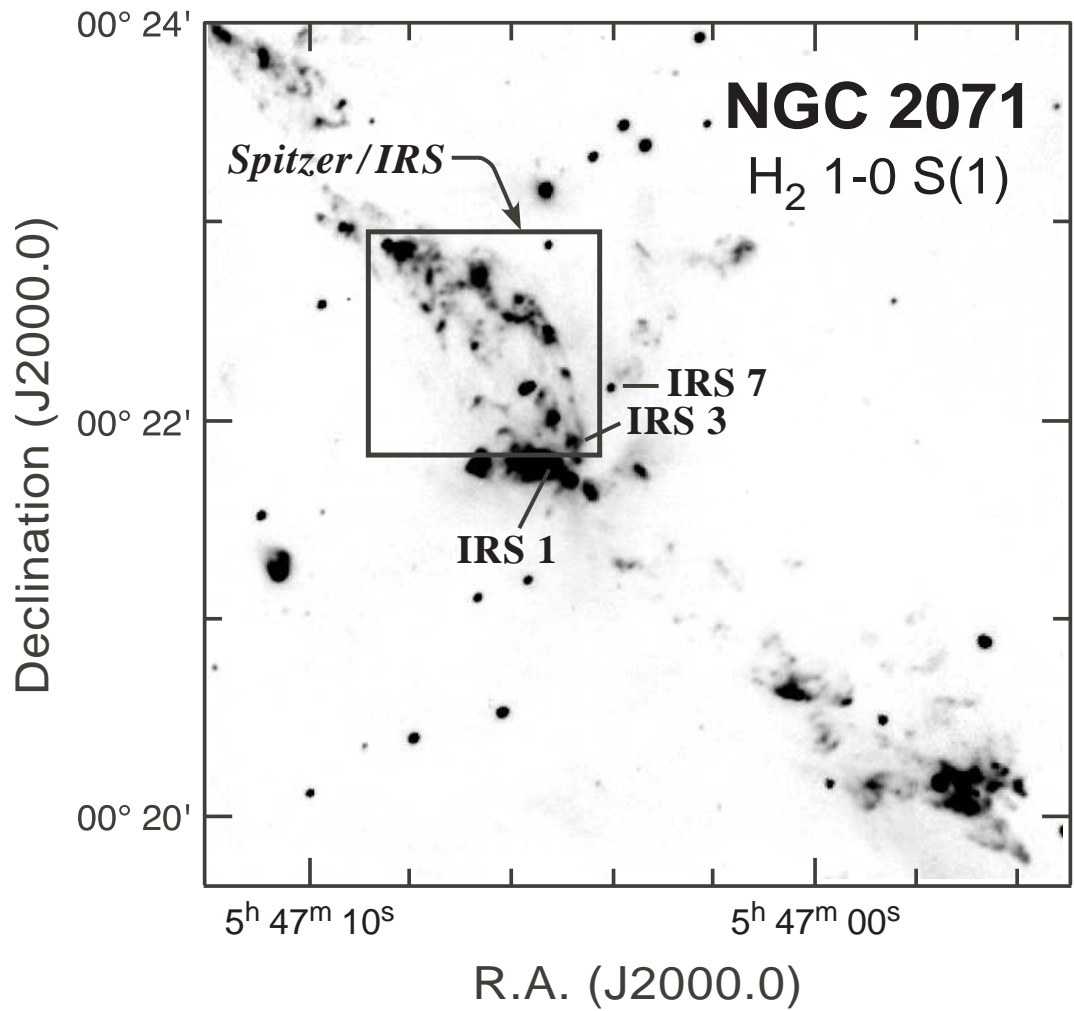}
\vspace{6mm}
\caption{Map of the northern lobe of the outflow from NGC 2071 as traced by the
H$_2$ 1-0 S(1) 2.12\um\ emission \citep[after][]{eisloffel00}.  The driving source of
the NGC 2071 outflow is believed centered close to  RA 05$^{\rm h}$ 47$^{\rm m}$ 
04$^{\rm s}\!\!.$6, Dec +00$^{\rm o}$ 21\amin\ 48\asec\ (J2000).
The box labeled {\em Spitzer/IRS} bounds the portion of the northern outflow lobe
observed here.}
\label{fig:Finder_Chart}
\end{figure}

NGC 2071 is a region of active star formation, located inside the northern
part of the Orion~B (L$\,$1630) molecular cloud complex at a distance of 390~pc
\citep{anthony82}.
The region of interest in this study lies approximately 4\amin\ north of the
optical reflection nebula NGC~2071 and is the site of a prominent outflow.  
Fig.~\ref{fig:Finder_Chart} shows a portion of the outflow, as delineated 
by the \mh\ 1-0 S(1) 2.12\um\ emission \citep{eisloffel00},
along with the region observed here.  Near-infrared 
continuum imaging and spectroscopy of this area reveal a cluster of infrared
sources near the center of this outflow \citep{walther91,walther93}.  
The detection of a bipolar thermal radio jet from source
IRS3 oriented in the direction of the NGC~2071 North flow, and of a ring of
\water\ maser spots perpendicular to the flow axis \citep{torrelles98},
suggest that IRS3 is the driving source of the NGC~2071 North flow.
The outflow, which extends more than 6\amin, has been mapped previously
in CO \citep{snell84, moriarty89}, CS \citep{zhou91},
SO, SiO, and HCO$^+$ \citep[e.g.,][]{chernin92, chernin93},
and in \mh\ 1-0 S(1) \citep[e.g.,][and references therein]{eisloffel00}.   
The presence of high-velocity gas -- as high as 60\kms\ -- and the 
enhanced abundance of species such as SO and SiO 
\citep[e.g.,][]{chernin93} suggest that
shock processing of the outflow gas is highly likely.

In this paper, we present the results of {\em Spitzer}/IRS observations 
toward the northern lobe of the NGC 2071 outflow and discuss the new
understanding revealed through high spatial resolution mapping in the
lines of H$_2$ S(0)$-$S(7), HD R(3) and R(4), Si$\,$II, Fe$\,$II, S$\,$I,
and three transitions of \nwater: 7$_{25}-$6$_{16}$, 6$_{34}-$5$_{05}$, and
7$_{34}-$6$_{25}$.  In addition, we present spectra that include several other
\water\ features along with several lines that we tentatively ascribe to OH.
Section~2 summarizes the observations and Section~3
presents the results.  Section~4 describes the calculations used to analyze
the line data and the conclusions are discussed in Section~5.

\section{\bf Observations and Data Reduction}

\vspace{-1mm}

NGC 2071 was observed on 20 and 21 October 2004 as part of {\em Spitzer
Space Telescope} Cycle 1 General Observer (GO) program 3423. 
The observations utilized the Long-High (LH), Short-High (SH),
and Short-Low (SL) modules to achieve the highest possible spectral
resolving power and complete spectral coverage available to IRS. 
The mapped areas were approximately 1\amin$\,\times\,$1\amin\ 
for LH and SH and
approximately 1\amin$\,\times\,$1.5\amin\  for SL, obtained by stepping in 
spacings of half-slit width perpendicular to the slit and in $\sim\,$4/5 
of the slit length parallel to the slit. The raw data were
processed with the Spitzer Science Center (SSC) pipeline 
version S12.0.2, which provided basic calibrated data (BCD). 
Further data reduction utilized a data pipeline developed
by our group.  This pipeline incorporates procedures for removing
bad or ``hot'' pixels, utilizes the SMART software package
\citep{higdon04} for calibration and extraction of spectra
at each position along the slit, applies a slit loss 
correction function (SLCF) to correct the flux calibration
performed for point-like sources to extended sources, and
automatically fits Gaussian lines and second-order polynomials
to a set of predetermined spectral lines at each position. The
first three steps of this data processing are described in
more detail in \citep{neufeld06a}.  The line fitting 
was performed iteratively to achieve the best-fit parameters. 
In some cases, the resulting line maps exhibited noticeable striping
attributable to varying pixel sensitivity across the slit and by the
after-effects of correcting for bad pixels.  
A pixel response correction factor (PRCF) 
was determined to remove striping in the final maps of all LH
and SH observations.  We used a statistical approach to determine the 
relative sensitivity in each pixel while preserving the total
flux in all pixels across the slit. Details of this 
method are provided in \citet{neufeld07}.

\section{\bf Results}

\vspace{-1mm}

\begin{figure}[t]
\centering
\includegraphics[scale=0.80]{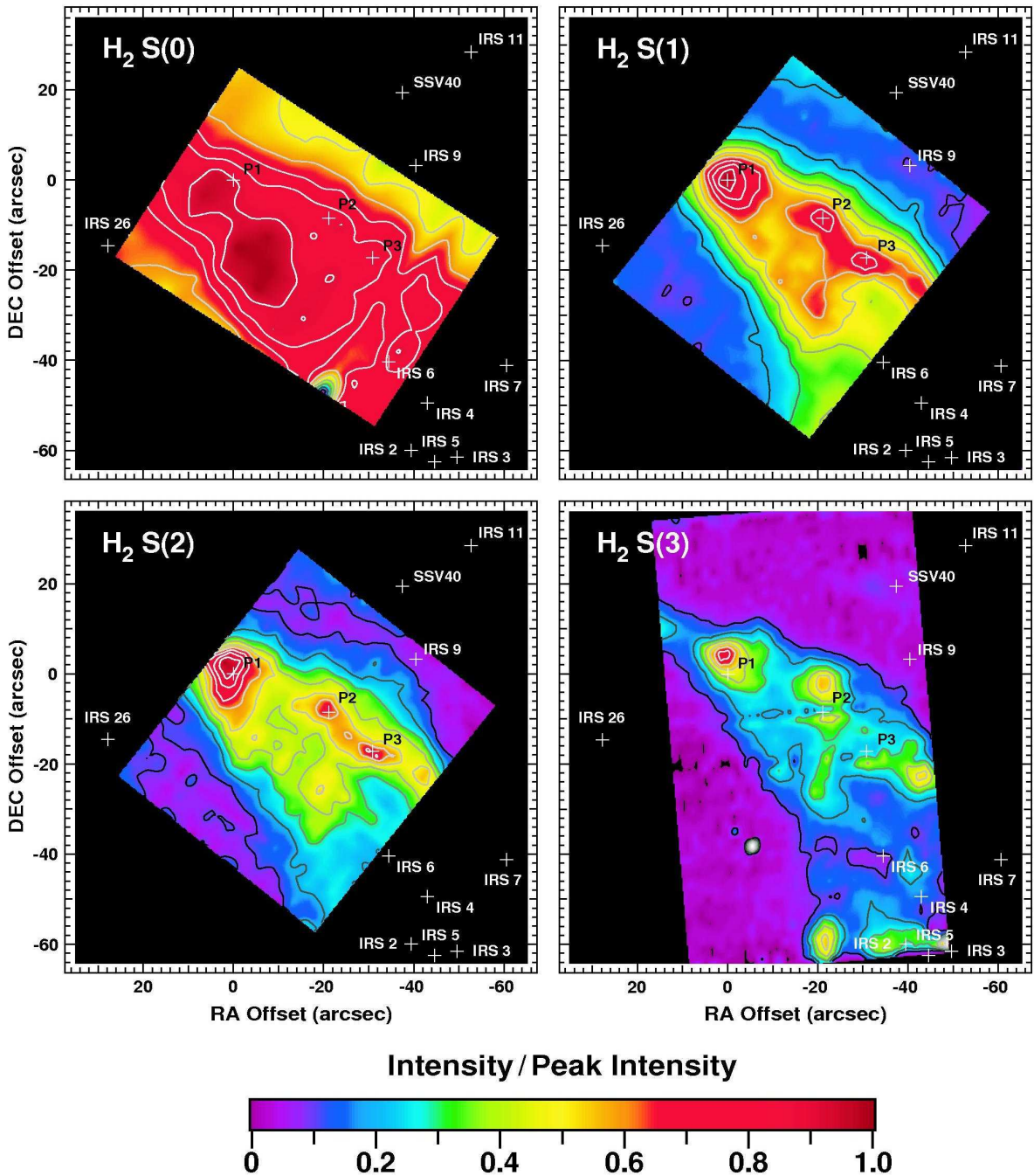}
\vspace{3mm}
\caption{Map of the northern lobe of the outflow from NGC 2071 as traced by
H$_2$ S(0) through S(3) emission.  Offsets are relative to the measured peak 
of the H$_2$ 17.035\um\ S(1) emission centered at RA 05$^{\rm h}$ 47$^{\rm m}$ 
08$^{\rm s}\!\!.$1, Dec +00$^{\rm o}$ 22\amin\ 50.7\asec\ (J2000).  The peak
intensity for each transition is provided in Table~1.  Positions P1, P2, and P3
are fiducials that mark the positions of the first, second, and third strongest
peaks in the H$_2$ S(1) emission.  The position of other known sources in
the field are also shown \citep[cf.][]{walther93}.}
\label{fig:map1}
\end{figure}

\begin{figure}[t]
\centering
\includegraphics[scale=0.80]{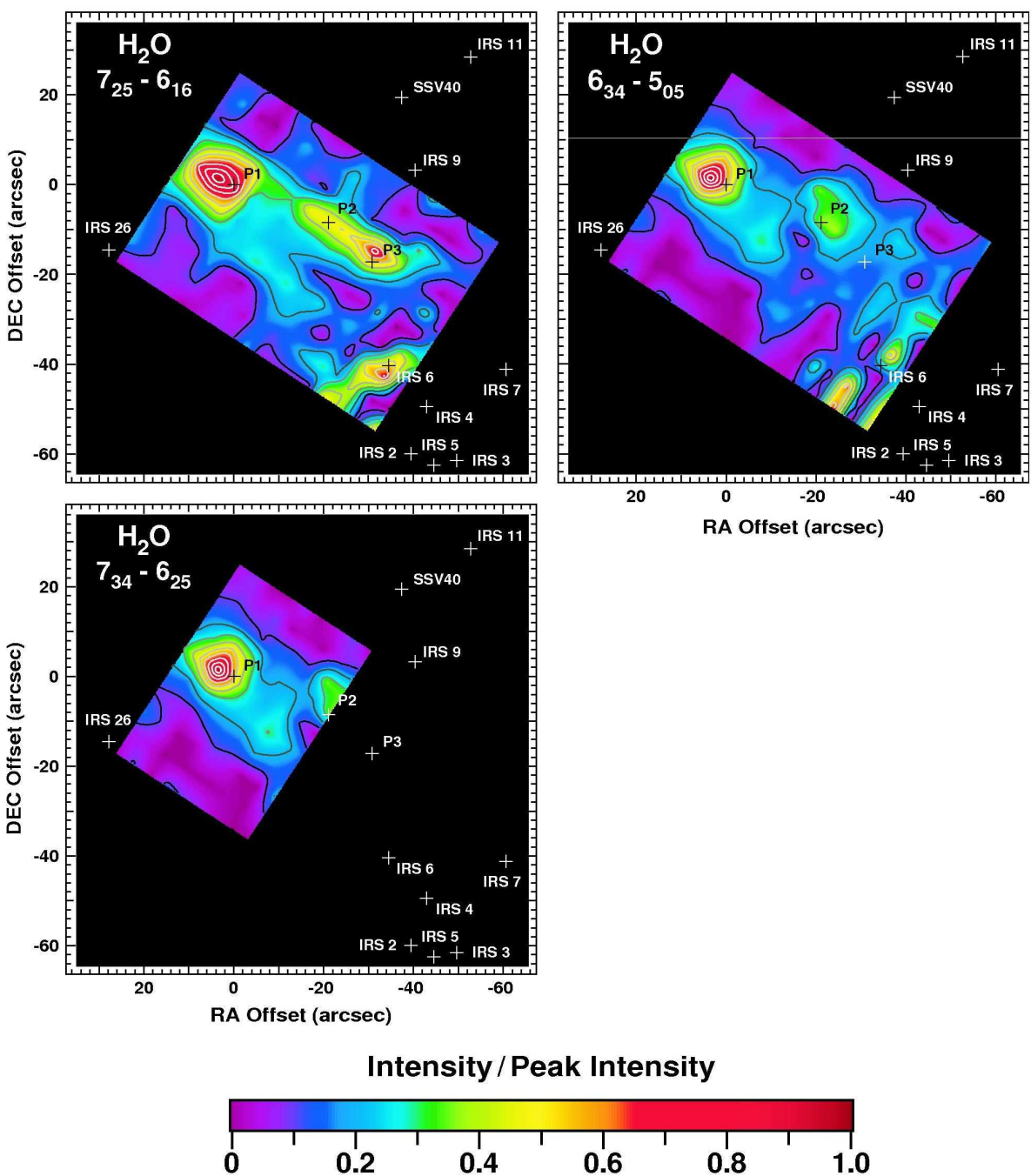}
\vspace{4mm}
\caption{Same as Fig.~2, except for H$_2$O 7$_{25}-$6$_{16}$ 29.837\um, 
6$_{34}-$5$_{05}$ 30.899\um\
(blended with H$_2$O 8$_{54}-$7$_{43}$ 30.871\um), and 7$_{34}-$6$_{25}$ 
34.549\um\ emission.  
The H$_2$O 7$_{34}-$6$_{25}$ map is truncated because {\em Spitzer}/IRS 
was exposed to a strong source while slewing to the NGC$\:$2071 field. 
This caused latent signal levels at the long wavelength end of LH and, thus, 
we conservatively omitted these less reliable observations.}
\label{fig:map3}
\end{figure}

\begin{figure}[b]
\centering
\includegraphics[scale=0.72]{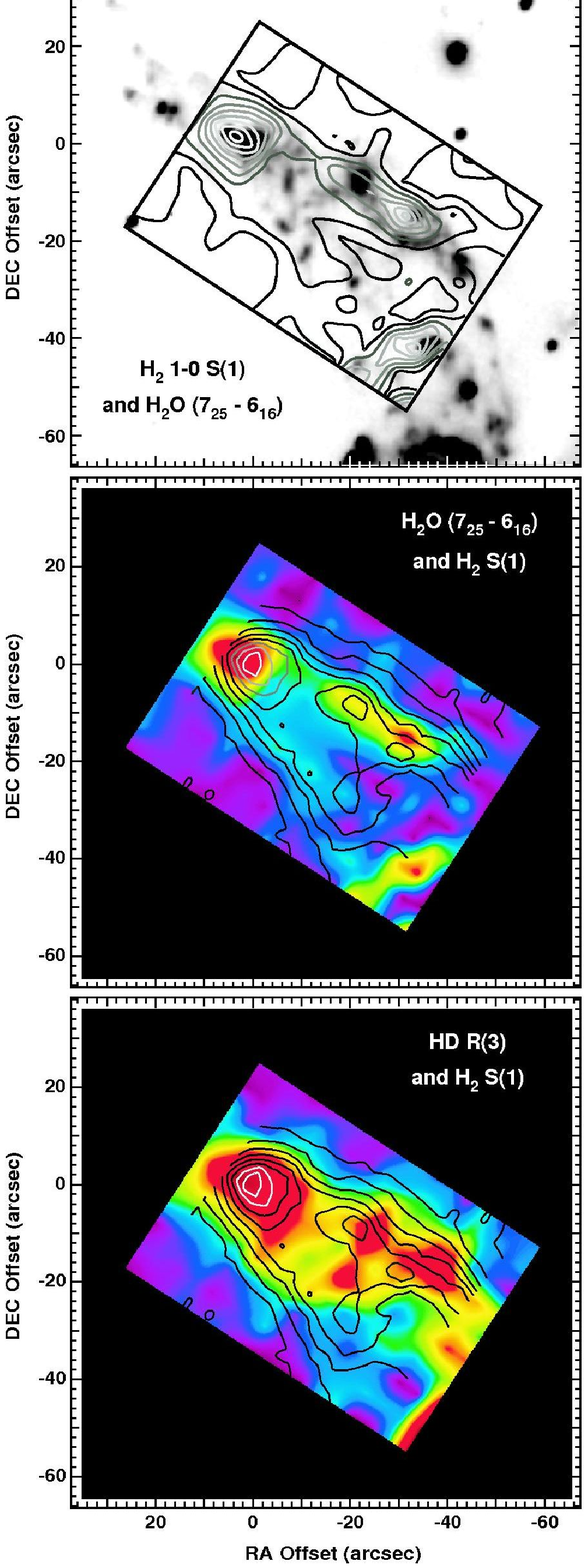}
\vspace{2mm}
\caption{{\em Top:} Contours of H$_2$O 7$_{25}-$6$_{16}$ 29.837\um\
emission superposed on the \citet{eisloffel00} H$_2$ 1-0 S(1) 2.12\um\ map.
{\em Middle:} Contours of H$_2$ S(1) emission superposed on the map of 
H$_2$O 7$_{25}-$6$_{16}$ 29.837\um\ emission. {\em Bottom:} 
Contours of H$_2$ S(1) emission superposed on the map of 
HD R(3) 28.502\um\ emission}
\label{fig:Overlay1}
\end{figure}

The observed spectra show emission from molecular, atomic and ionic species
as well as a number of PAH features.  In particular, we detect emission from
[S$\,$I], [S$\,$III], [Fe$\,$II] and [Si$\,$II] transitions, H$_2$ rotational 
lines from S(0) to S(7), HD R(3) and R(4) transitions, and at least 11 H$_2$O
transitions.  Most of the line emission was strong enough to detect over
extended regions and these maps are shown in 
Figs.~\ref{fig:map1}$\,$-$\,$\ref{fig:map4}.  All maps have
been normalized to the peak line intensities listed
in Table~1.  For reference, the positions of the three strongest
peaks in the H$_2$ S(1) 17\um\ emission are marked as P1, P2,
and P3 on these maps.
As can be seen in Fig.~\ref{fig:map3}, the strongest water emission
lies close to the peak of the H$_2$ S(1) 17\um\ emission (position
P1 in Fig.~\ref{fig:map3}).  It is also clear that: (1) the water emission is
extended; (2) this extended emission is generally aligned with
the outflow; and, (3) the water emission exhibits secondary peaks 
toward positions of enhanced H$_2$ emission.
Fig.~\ref{fig:Overlay1} shows a direct comparison of the spatial distribution
of the H$_2$O (7$_{25}\,-\,$6$_{16}$) emission versus that due to
the ro-vibrational
H$_2$ 1-0 S(1) and pure rotational H$_2$ S(1) emission.
While the overall distribution of the two species is quite similar, the peaks
of the H$_2$O and H$_2$ are sometimes offset with respect to each other.
For example, the peaks of the H$_2$O 7$_{25}\,-\,$6$_{16}$ and the
H$_2$ S(1) emission are offset by 3.8\asec, which exceeds the stated
IRS sky registration error, and thus is likely real.

\begin{figure}[b]
\centering
\includegraphics[scale=0.76]{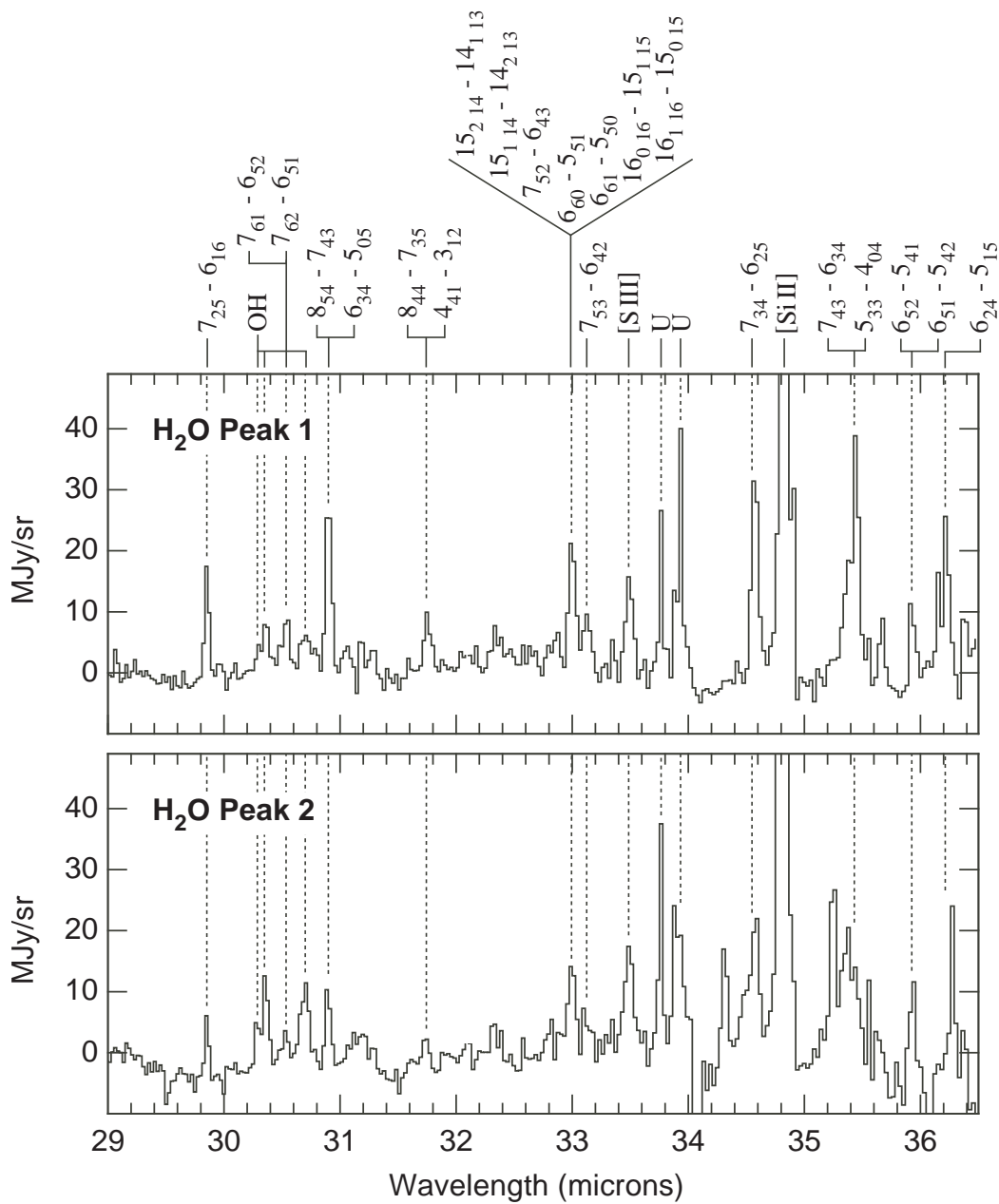}
\vspace{3mm}
\caption{Portion of the {\em Spitzer/IRS} spectra obtained in a 15\asec\
synthesized beam toward H$_2$O 
Peaks 1 and 2 (see text).  The \nwater\ transition(s) associated with each
detected feature are shown.  In the absence of higher spectral resolving
power, we are not able to observationally determine the extent to which
each transition contributes to the total flux measured for each blended
feature.  In addition to Si$\,$II and S$\,$III lines in the band, we
also note the tentative detection of OH $^2\Pi_{3/2}$ 19/2$^+$-17/2$^-$
30.277\um, $^2\Pi_{3/2}$ 19/2$^-$-17/2$^+$ 30.346\um,
$^2\Pi_{1/2}$ 17/2$^+$-15/2$^-$ 30.657\um.  Unidentified lines are
denoted with a ``U".}
\label{fig:Spectra1}
\end{figure}

\begin{figure}[b]
\centering
\includegraphics[scale=0.76]{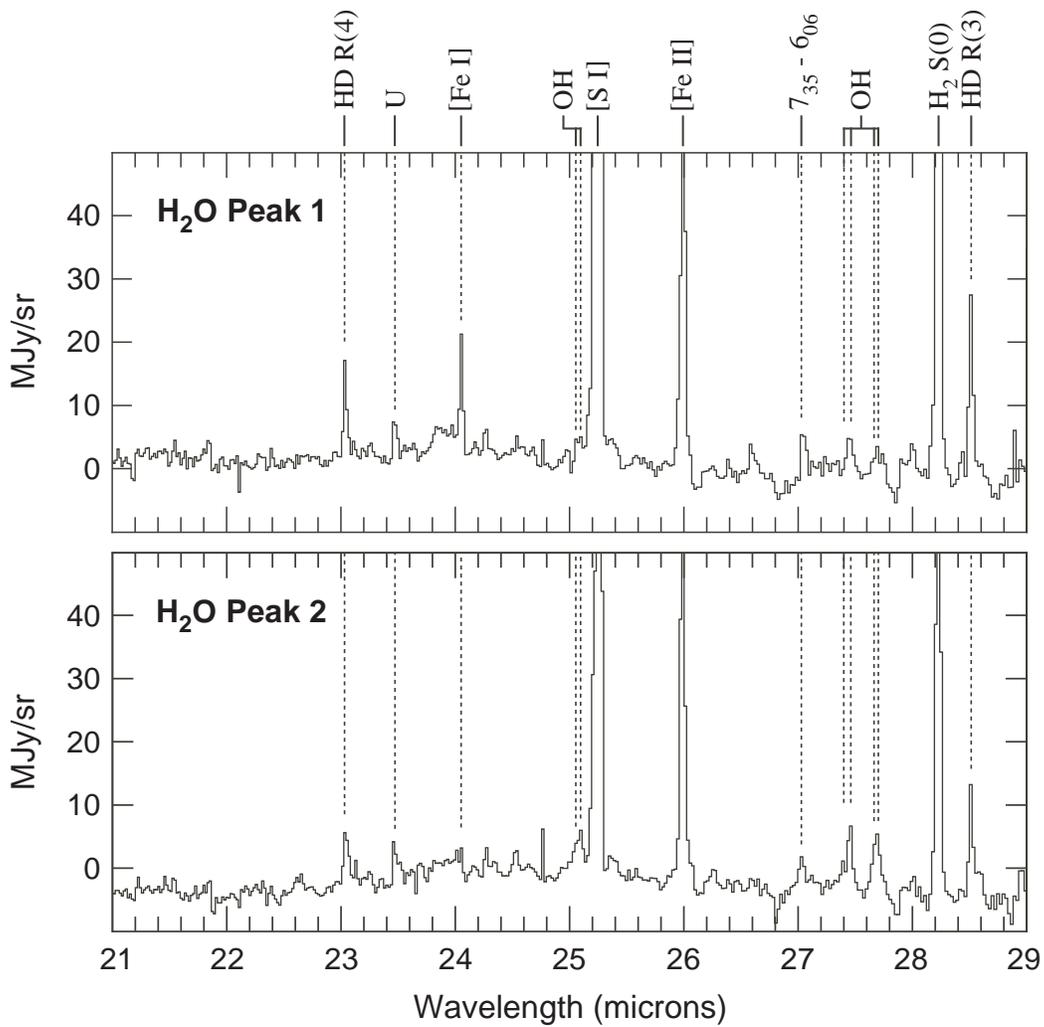}
\vspace{6mm}
\caption{Portion of the {\em Spitzer/IRS} spectra obtained in a 15\asec\
synthesized beam toward H$_2$O 
Peaks 1 and 2 (see text).   Of particular note is the ground rotational
transition of para-H$_2$ at 28.219\um, two HD rotational transitions,
R(3) and R(4), at 28.502 and 23.034\um, respectively, and the tentative detection 
of OH $^2\Pi_{3/2}$ 21/2$^-$ - 19/2$^+$
27.393\um, $^2\Pi_{3/2}$ 21/2$^+$ - 19/2$^-$ 27.454\um, 
$^2\Pi_{1/2}$ 19/2$^-$ - 17/2$^+$ 27.652\um,
$^2\Pi_{1/2}$ 19/2$^+$ - 17/2$^-$ 27.697\um,
$^2\Pi_{3/2}$ 23/2$^+$ - 21/2$^-$ 25.035\um, and
$^2\Pi_{3/2}$ 23/2$^-$ - 21/2$^+$ 25.090\um.  Unidentified lines are
denoted with a ``U".}
\label{fig:Spectra2}
\end{figure}

\begin{figure}[b]
\centering
\includegraphics[scale=0.76]{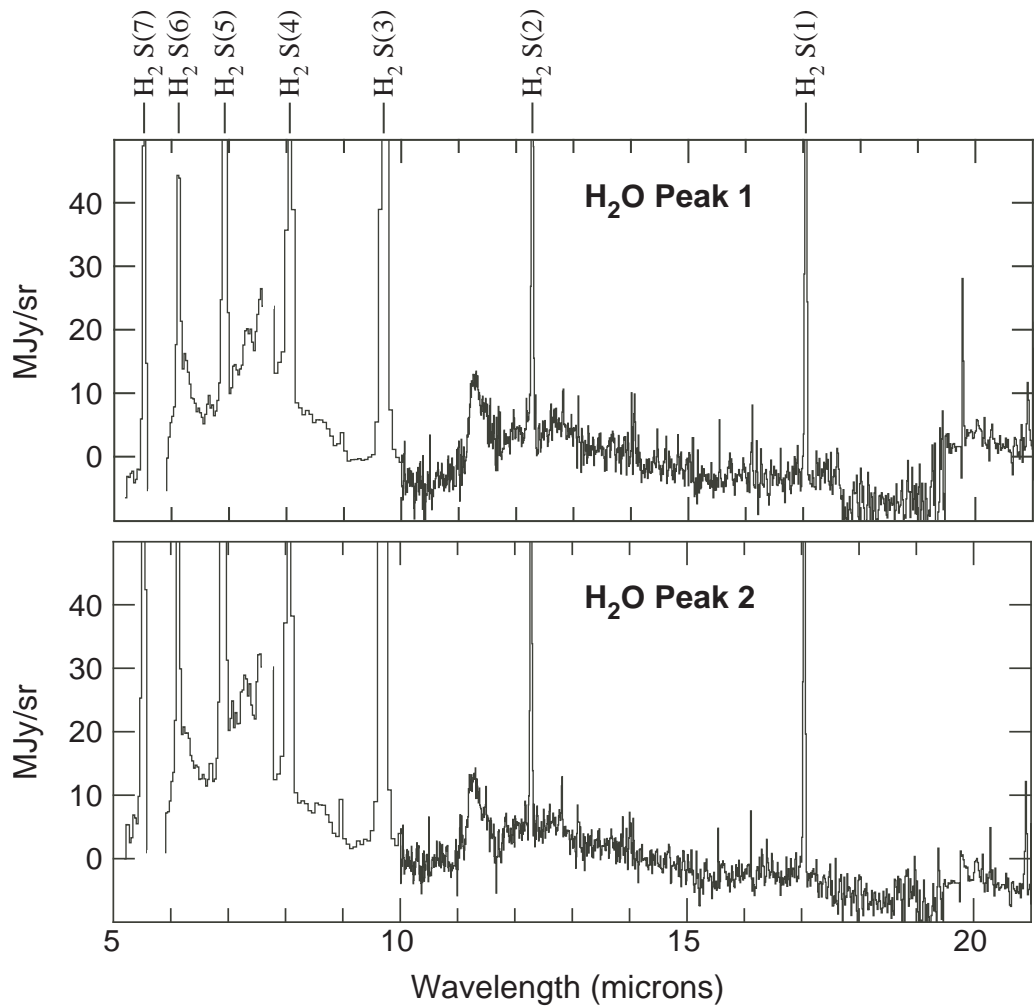}
\vspace{6mm}
\caption{Portion of the {\em Spitzer/IRS} spectra obtained in a 15\asec\
synthesized beam toward H$_2$O 
Peaks 1 and 2 (see text).  Of note are the prominent lines of H$_2$ S(1) through
(7).  Both spectra also show broad PAH emission features with peak 
wavelengths of 6.22, 7.41, 7.59, 7.85, 8.61, and 11.3\um.}
\label{fig:Spectra3}
\end{figure}

The spectra obtained toward the water emission peaks closest to
positions P1 and P2 are shown
in Figs.~\ref{fig:Spectra1}$\,-\,$\ref{fig:Spectra3}.
In order to reduce the baseline noise, we averaged the data 
toward the water peaks using a Gaussian weighing function with a 
wavelength-dependent HPBW of between 11.0\asec\ at 27.0\um\ and 
14.7\asec\ at 5.5\um.  The final synthetic beams have effective beam 
widths of 15\asec\ for all wavelengths of interest \citep[cf.][]{neufeld06a}.
The spectrum labeled H$_2$O Peak~1 is centered at
RA 05$^{\rm h}$ 47$^{\rm m}$ 
04$^{\rm s}\!\!.$6, Dec +00$^{\rm o}$ 22\amin\ 52.2\asec\ (J2000),
while the spectrum labeled H$_2$O Peak~2 is centered at
RA 05$^{\rm h}$ 46$^{\rm m}$ 
40$^{\rm s}\!\!.$4, Dec +00$^{\rm o}$ 22\amin\ 43.0\asec\ (J2000).
For almost all of the H$_2$O and H$_2$ transitions detected, the 15\asec\
synthetic beam is sufficiently large to encompass the water emission
as well as the bulk of the H$_2$ emission toward each peak.
The measured line fluxes for the water features associated with 
H$_2$O Peak~1 are provided in Table~2.  Unfortunately, at the spectral
resolving power of the {\em Spitzer}/IRS, many of the \water\ features
are blends of two or more \water\ transitions.  The extinction-corrected
line fluxes for
the other key species in this spectrum are given in Table~3.
We also note the tentative detection of emission from a number of
OH rotational transitions (see Figs.~\ref{fig:Spectra1} and \ref{fig:Spectra2})
whose upper levels lie as much as $\sim\,$3500~K above the ground.

Direct measures of the extinction toward the NGC 2071 outflow are
not available.  Instead, we inferred the required extinction corrections
as a function of wavelength in 
the {\em Spitzer} bandpass by fitting the wings  of the silicate absorption 
feature at 9.7\um, hence minimizing the effects of line saturation 
when present.  Our silicate fitting model uses the renormalized 
synthetic Galactic extinction curve per unit hydrogen column density 
calculated for \rv$\,=\,$3.1 or for \rv$\,=\,$5.5 
\citep{draine03a, draine03b} in 
linear combination with a first-degree polynomial and 4 Gaussians with 
fixed widths and fixed centers to account for emission from interfering 
lines.  Since the gas clumps probed by our data have dust properties most 
probably intermediate between diffuse and dense molecular clouds, the 
extinction corrections we adopted represent the mean of the 
extinction corrections we obtained for \rv$\,=\,$3.1 and those we 
obtained for \rv$\,=\,$5.5.

\section{\bf Water Calculations}

\vspace{-1mm}

To assess how the \water\ line strengths
measured here constrain the water abundance, the equilibrium level
populations of all ortho and para rotational levels of the ground
H$_2^{\;16}$O vibrational state with energies {\em E/k} up to 7700~K
have been calculated using an escape probability method described by
\citet{neufeld91}.  
In modeling the excitation of water vapor, we are limited by the 
availability of molecular data.  Thus far, at the high temperatures of 
relevance here, a complete set of rate coefficients has only been 
computed for collisional excitation of H$_2$O by He \citep{green93}.
A new and accurate nine-dimensional potential energy 
surface has recently been computed for H$_2$O-H$_2$ \citep{faure05},
and quantal calculations are currently underway 
\citep[][personal communication]{dubernet07},
to obtain rate coefficients for the collisional 
excitation of H$_2$O by H$_2$.  However, quantal results for transitions 
involving the highly-lying rotational states observed by {\em Spitzer} have 
not yet been completed.   Nevertheless, for the subset of those 
transitions that show the largest rate coefficients, \citet{faure07}
have recently performed quasi-classical trajectory (QCT) 
calculations to estimate the collisional rates, and have kindly made the 
results available to us prior to publication.  Unfortunately, 
transitions with small rate coefficients still play a significant role 
in the excitation of the spectral lines that we have observed with 
{\em Spitzer}; for those transitions, we have had to use the \citet{green93}
results, scaled by a factor 1.348 to account for the difference 
between the H$_2$O-He and H$_2$O-H$_2$ reduced masses.
Finally, a polynomial fit to the exact expression for the photon escape
probability from a plane-parallel emitting region \citep{hummer82}
is used.

We assume that the physical conditions within H$_2$O Peak 1 are
those required to reproduce the measured H$_2$ S(0)
through S(7) flux from the same region.  Table~4 summarizes
these physical conditions.
The model H$_2$ column densities are assumed to 
be those consistent with a simple shock model in which

\begin{equation}
N(H_2)~=~6.45 \times\ 10^{20}\;\left[\frac{n(H_2)}{10^5}\right]^{0.5}\;
\left[\frac{T_{gas}}{1000}\right]^{-0.555}~~~{\rm cm^{-2}},
\end{equation}

\noindent where $n(H_2)$ is the density in the postshock gas
\citep[cf.][]{neufeld06a}.
We assume that the fraction of the beam filled
by the warm and hot H$_2$O is given by the ratio of the measured
values of N(H$_2$) given in Table 4 to those derived above.
To estimate the physical parameters from the eight \mh\ rotational
lines, we have performed calculations in which the emitting region
is approximated by a two-component model for two 
different cases: (1) clouds with homogenous density; and, (2) clouds with 
homogenous pressure. The six best-fit parameters (temperature, column 
density and \mh\ ortho-to-para ratio for the two components) are 
determined by minimizing the root mean square of ln$\,$(measured 
flux/computed flux) for the \mh\ S(0) to S(7) lines. Results were 
obtained for a series of H$_2$ number densities ranging from 
10$^4$ to 10$^7$ \cmc, and pressures ($\times\,$1$/k$) 
ranging from 10$^7$ to 10$^{10}$ \cmc$\:$K.  The greatest differences 
between the two models -- homogenous density or pressure -- appear at 
low densities (or low pressures), where the pressure-balanced 
assumption requires a relatively higher temperature for the hot component. 
The results of the two models converge as LTE is approached
for the higher-lying transitions, i.e., for number densities greater 
than 10$^5$ \cmc\ (or pressures/$k$ greater than 10$^8$ \cmc$\;$K). 
In this case, the best-fit temperature for the warm component is around
300-360~K while that for the hot component is around 1000-1500~K.

Using the physical conditions that best fit the \mh\ observations, the  
abundances of H$_2$O and HD toward H$_2$O Peak~1 were then 
derived.  For \water, the total (ortho+para) abundance at each assumed density 
was varied to produce the best overall fit to the features listed in Table~2.  We choose
these features to fit because they are the strongest observed transitions between
5 and 36.5\um\ toward H$_2$O Peak 1; both the observed and predicted line 
fluxes for \water\ transitions lying at $\lambda <\,$29\um\ are less than 
5~MJy/sr, and most typically less than about 3 MJy/sr, rendering their
unambiguous identification above the noise difficult.  As discussed further below,
the \water\ ortho-to-para ratio was treated as a free parameter.
Similarly, the abundance of HD was determined by fitting the measured
R(3) and R(4) line fluxes, although the best-fit HD abundance is not a 
sensitive function of number density 
due to the moderately low critical densities of the R(3) and R(4) transitions
of $\sim\,$10$^4$ to 10$^5$ \cmc.  Finally, it would be useful to also fit the
measured OH rotational lines.  Unfortunately, the lack of reliable collisional rate 
coefficients for these high-lying transitions makes the results of such a calculation
uncertain at this time.

As shown in Fig.~\ref{fig:Bestfitdensity}, fits to the \mh\ and HD emission
lines alone favor gas with higher densities or pressures.   
The best-fit HD abundance relative to \mh\ is between 1.1$\times$10$^{-5}$ and 
1.8$\times$10$^{-5}$, with only a weak dependence on density as noted above.
This abundance is very similar to those derived previously from observations of 
HD R(3) and R(4) in other shocked regions that we have observed with
{\em Spitzer} \citep{neufeld06b}.
Fits to the \water\ features were computed in two ways.  First, fits to ten of the
eleven \water\ features in Table~2 were calculated assuming an \water\ ortho-to-para
ratio of 3:1 as well as an ortho-to-para ratio that was allowed to vary in order to
achieve the best overall fit at each density between 10$^4$ and 10$^7$~\cmc.  
The strong
feature at 34.55\um\ was not included.  The conditions needed to reproduce 
the flux in this line resulted in fluxes in the other \water\ transitions that are in 
significant disagreement with the observations.  Thus, we conclude that either this 
line is blended with an unknown feature or uncertainties in the collision rates may be 
resulting in lower-than-actual populations in the upper 7$_{34}$ state.
Second, to better isolate the effects of varying the \water\ ortho-to-para ratio, we
examined the three strongest features in which the only contribution to the line flux
was an ortho-\water\ transition, at 29.84\um, or a para-\water\ transition, at
33.13 and 36.21\um.  These results are also shown in Fig.~\ref{fig:Bestfitdensity}.
Based on fits to the water data alone, \mh\ densities in the water-emitting region of
between about 3$\times$10$^4$ and 3$\times$10$^5$~\cmc\ appear favored.
The best fits to the isolated ortho and para features are achieved for \mh\ densities
between 3$\times$10$^4$ and 10$^5$~\cmc\ and an assumed \water\ ortho-to-para
ratio of between 1.5 and 1.6.

\begin{figure}[b]
\centering
\includegraphics[scale=0.77]{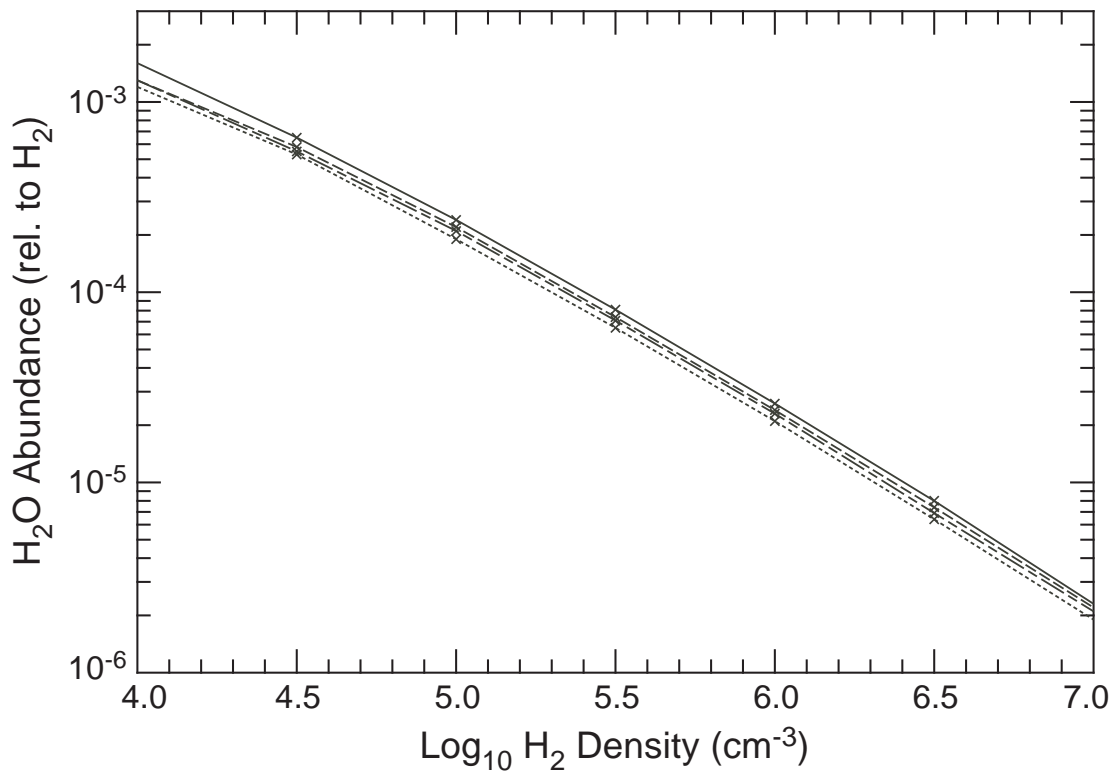}
\vspace{8mm}
\caption{Total (ortho+para) water abundance, at each assumed density, that produces
the best overall fit to: (1) the \water\ features listed in Table~2, with the
exception of the 34.55\um\ feature, and assuming an ortho-to-para
ratio of 3:1 ({\em dashed line}); 
(2) the \water\ features listed in Table~2, with the
exception of the 34.55\um\ feature, and assuming an ortho-to-para
ratio that has been allowed to vary to produce the best fit to the
data at each density ({\em dot-dashed line}); 
(3) the 29.84, 33.13, and 36.21\um\ features and assuming an ortho-to-para
ratio of 3:1 ({\em solid line}); and,
(4) the 29.84, 33.13, and 36.21\um\ features and assuming an ortho-to-para
ratio that has been allowed to vary to produce the best fit to the
data at each density ({\em dotted line}).}
\label{fig:Bestfitabundance}
\end{figure}

The best-fit water abundances are shown in Fig.~\ref{fig:Bestfitabundance}.
To a good approximation, the best fits to the \water\ line fluxes are proportional
to the \mh\ density 
$\times$ the water abundance, resulting in a derived number density of 
\water\ molecules of 12$\,-\,$29~\cmc, independent of \mh\ density.
This is a consequence of the fact that the observed transitions  
have critical densities $>\,$10$^9$~\cmc\ and are optically thin under the
conditions considered here.
In this limit, the flux in an ortho-\water\ (or para-\water)  line is

\vspace{-3mm}

\begin{equation}
F = \frac{A_{u\ell}\,h\nu_{u\ell}\,N_u\,\Omega}{4\pi} = 
\frac{A_{u\ell}\,h\nu_{u\ell}\,\left[n({\rm H_2}) \cdot \ell \cdot \chi({\rm H_2O}) 
\cdot f_u \cdot \xi_{\rm op} \right]\,\Omega}{4\pi}~~,
\end{equation}

\vspace{3mm}

\noindent where $A_{u\ell}$ is the transition spontaneous emission rate,
$\nu_{u\ell}$ is the transition frequency,
$n$(H$_2$) is the H$_2$ density, $\ell$ is the depth of
the emitting region, $\chi$(H$_2$O) is the water abundance
relative to H$_2$, $f_u$ is the
fractional population in the upper state of the emitting transition, 
$\xi_{\rm op}$ is the fraction of all \water\ molecules in the ortho (or para)
state, and $\Omega$ is the solid angle of 
the water emitting region.

\begin{figure}[b]
\centering
\includegraphics[scale=0.81]{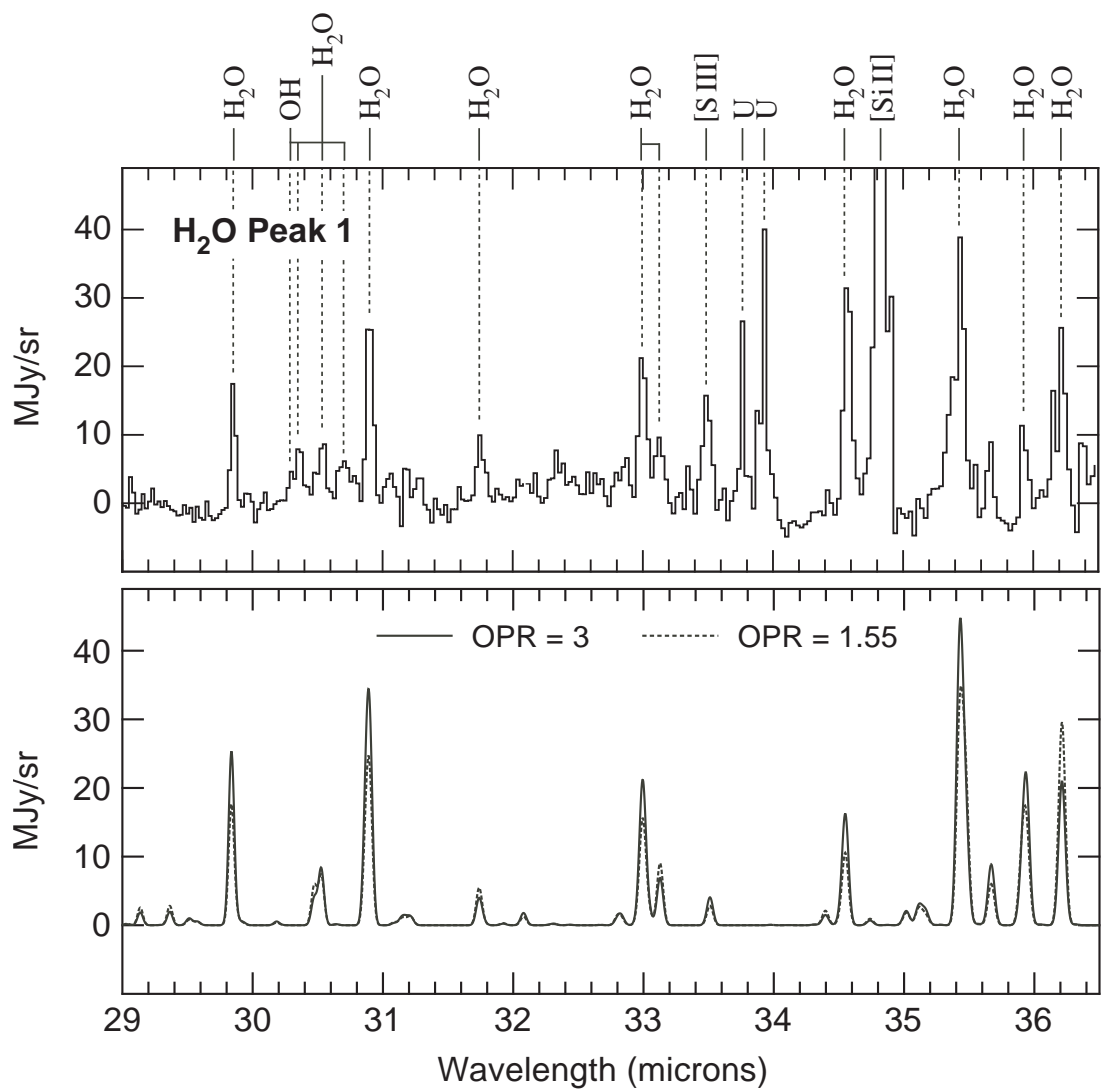}
\vspace{8mm}
\caption{{\em Top}: Observed spectrum within 15\asec\ synthesized beam
toward H$_2$O Peak 1.  {\em Bottom}: Model spectra assuming 
$n$(H$_2$) = 10$^5$~\cmc\ and an \water\ ortho-to-para ratio of 3:1
({\em solid line}) and an \water\ ortho-to-para of 1.55:1 ({\em dotted line}),
which provides the best fit to the pure ortho 29.84\um\ and pure
para 33.13 and 36.21\um\ lines at this density.}
\label{fig:Computedspectra}
\end{figure}

Fig.~\ref{fig:Computedspectra} shows the resulting 29$\,-\,$36.5\um\ water 
spectrum computed for an \mh\ density of 10$^5$~\cmc\ and \water\
ortho-to-para ratios of 3:1 and 1.55:1, the latter being the best-fit ortho-to-para
ratio at a density of 10$^5$~\cmc.
The computed spectra were convolved with a Gaussian profile
whose width was selected to match the IRS Long-High resolving power,
$\lambda/\Delta\lambda$, of 600 (FWHM).  As can be seen in 
Fig.~\ref{fig:Computedspectra}, the differences between the
spectrum computed assuming an \water\ ortho-to-para ratio of 3:1 and 1.55:1
are small.  Nevertheless, assuming an ortho-to-para ratio less than 3:1 results in
a slightly better overall fit to the \water\ data and suggests a possible departure
from the usual assumption that water in post-shocked gas is always 
present in the LTE ortho-to-para ratio of 3:1.  This tentative finding is discussed
further in the next section.

Taking into account the \mh, HD, and \water\ spectra, the overall 
best fits to these data favor an H$_2$ density in the range 3$\times$10$^4$ to 
10$^6$~\cmc, and 
2$\times$10$^{-5}\;\simlt\;\chi($H$_2$O)$\;\simlt\;$6$\times$10$^{-4}$.
Unfortunately, the present data do not allow us to better constrain the
H$_2$ density, and thus the water abundance, within the emitting region.
However, as we discuss in the next section, future observations planned for
the {\em Herschel Space Observatory} may permit more direct measures of 
the \mh\ density.

\section{\bf Discussion}

\vspace{-1mm}

The capabilities of the {\em Spitzer}/IRS instrument has enabled the study
of outflow regions in ways not possible previously.  In particular, the
ability to map the distribution and intensity of key species, like
H$_2$, HD, Fe$^+$, Si$^+$, S, and, of course, H$_2$O, with a spatial resolution
of a few arcseconds permits more stringent tests of  our understanding of 
the structure and composition of outflows.  Earlier space missions,
such as the {\em Infrared Space Observatory} (ISO), {\em SWAS}, and 
{\em Odin} have established the correlation between outflows from
young stellar objects and strong water emission 
\citep[e.g.,][]{harwit98, melnick00, neufeld00, nisini00, 
giannini01, benedettini02}.  
Unfortunately, these observations were conducted
with large beams relative to
{\em Spitzer}/IRS or, in the case of the {\em ISO} Short-Wavelength
Spectrometer with its 10\asec\ beam, lower sensitivity than {\em Spitzer}/IRS.  
As a result, the
filling factor of \water\ within the beam was not always directly known
or key species helpful in the interpretation of the water data, like H$_2$ 
and HD, were not detected.  The observations reported here remedy most,
but not all, of these previous shortcomings.

The primary goal of this study is the further understanding of water in outflows.
Unlike the production of water in quiescent molecular clouds, which proceeds
via a set of relatively slow ion-neutral reactions in the gas phase 
\citep[e.g.,][]{herbst73}, or through the photodesorption of water formed
on the surface of dust grains \citep[e.g.,][]{hollenbach08}, the material in
outflows often passes through a shock which both compresses and heats the
gas.  In the post-shocked region of non-dissociative, or $C$-type, shocks
\citep{draine80}, with shock velocities, $v_s$, $\simgt\,$10\kms, temperatures
exceed 300~K enabling a pair of neutral-neutral reactions
(H$_2$ + O $\rightarrow$ OH + H and 
H$_2$ + OH $\rightarrow$ H$_2$O + H) that can rapidly convert
all gas-phase oxygen not bound in CO into \water\ \citep{eanddj78}.
For $v_s\,\simgt\,$50\kms, most $C$-type shocks break down, giving rise
to dissociative, or $J$-type, shocks.
Molecules are completely dissociated, either by the precursor UV field generated
by the hot gas near the shock front or by collisions in the shock, and then
reform on grain surfaces in the cooler ($\sim\,$400~K) downstream gas
\citep{hollenbach89}.  Because $C$-type shocks have very low ionic 
abundances, strong Si$\,$II (34.8\um) and Fe$\,$II (26.0\um) emission
is considered a tracer of $J$-type shocks.

\begin{figure}[t]
\centering
\includegraphics[scale=0.80]{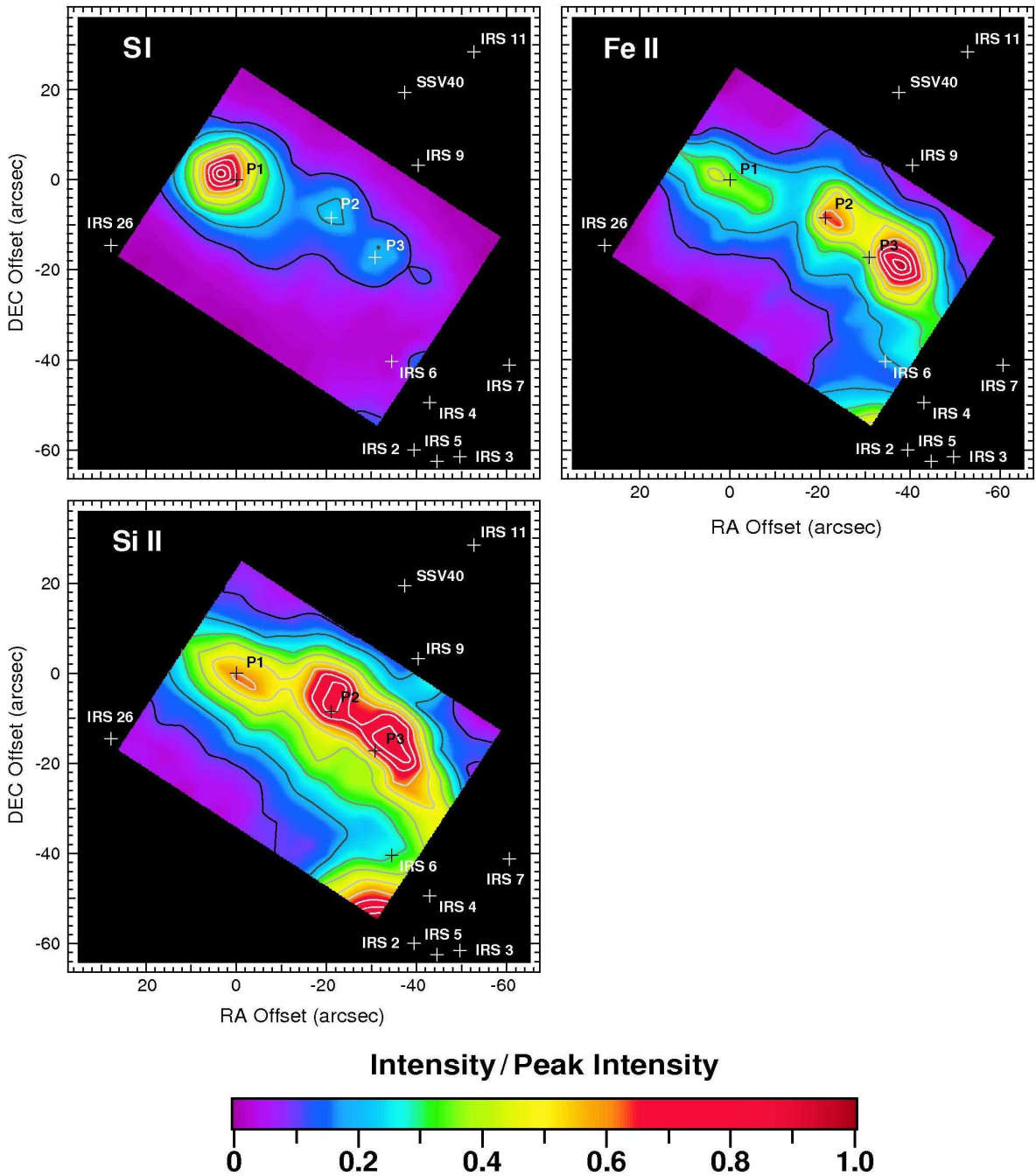}
\vspace{4mm}
\caption{Same as Fig.~2, except for 
SI $^3$P$_1-^3$P$_2$ 25.249\um, Fe$\,$II $a^6$D$_{7/2}-a^6$D$_{9/2}$
25.988\um, and Si$\,$II $^2$P$_{3/2}-^2$P$_{1/2}$ 34.815\um\ emission.}
\label{fig:map4}
\end{figure}

The {\em Spitzer}/IRS data suggest that both $J$- and $C$-type shocks are present
in the portion of the outflow observed here.  The strong Si$\,$II and Fe$\,$II
emission evident in Fig.~\ref{fig:map4} follows the outflow, peaking close to
H$_2$ positions P3 and P2, and lies closer to the source of the
outflow than the most prominent H$_2$ position, P1.  This may indicate the 
presence of fast-moving gas along the center of the outflow which has been 
slowed as it approaches position P1.  This picture is supported by previous
CO $J =\,$3$-$2 observations \citep{chernin92}, which
revealed well collimated, high velocity (30$\,-\,$60\kms) gas in a 
symmetric bipolar outflow only within 1.5\amin\ of the source, presumed
close to IRS$\:$3.  Moreover,
the CO data show velocity peaks appearing at $\pm\,$1\amin\ from the 
source.  Further support is provided by the
pronounced peak in the S$\,$I (25.2\um) emission (see Fig.~\ref{fig:map4})
near position P1.  In the $J$-type shock models of \citet{hollenbach89}, 
for densities between 10$^4$ and 10$^6$~\cmc, the strongest S$\,$I 
emission arises at the lowest velocities (i.e., $v_s\,\sim\,$30\kms), whereas 
both the Si$\,$II and Fe$\,$II emission increase with shock velocity.  
In addition, previous {\em Spitzer}/IRS observations of supernovae remnants
\citep{neufeld07} show a strong correlation between the S$\,$I and
H$_2$ S(1)$-$S(7) spatial distribution, suggesting that most of the S$\,$I 
emission arises in the slower, nondissociative shocks.  This could account for
the difference in the spatial distribution of Si$\,$II, Fe$\,$II, and S$\,$I.

That the H$_2$ S(1) peak emission appears to be located closer to the 
source of the outflow (near IRS~3) than all of the higher-lying H$_2$
transitions, i.e. S(2) through S(7), as well as that of the H$_2$O may also
be explained by this scenario.  In the multiple shock picture, the H$_2$ S(1)
emission could arise primarily in the $\sim\,$400~K gas behind a moderate
$J$-type shock -- the post shock region in which it is predicted that H$_2$
is reformed \citep{hollenbach89}.  Meanwhile, the higher-lying
H$_2$ and H$_2$O lines would arise from behind the $C$-type shock
slightly farther from the outflow source in which gas temperatures
can heat molecules to more than 2000~K.

\begin{figure}[b]
\centering
\includegraphics[scale=0.77]{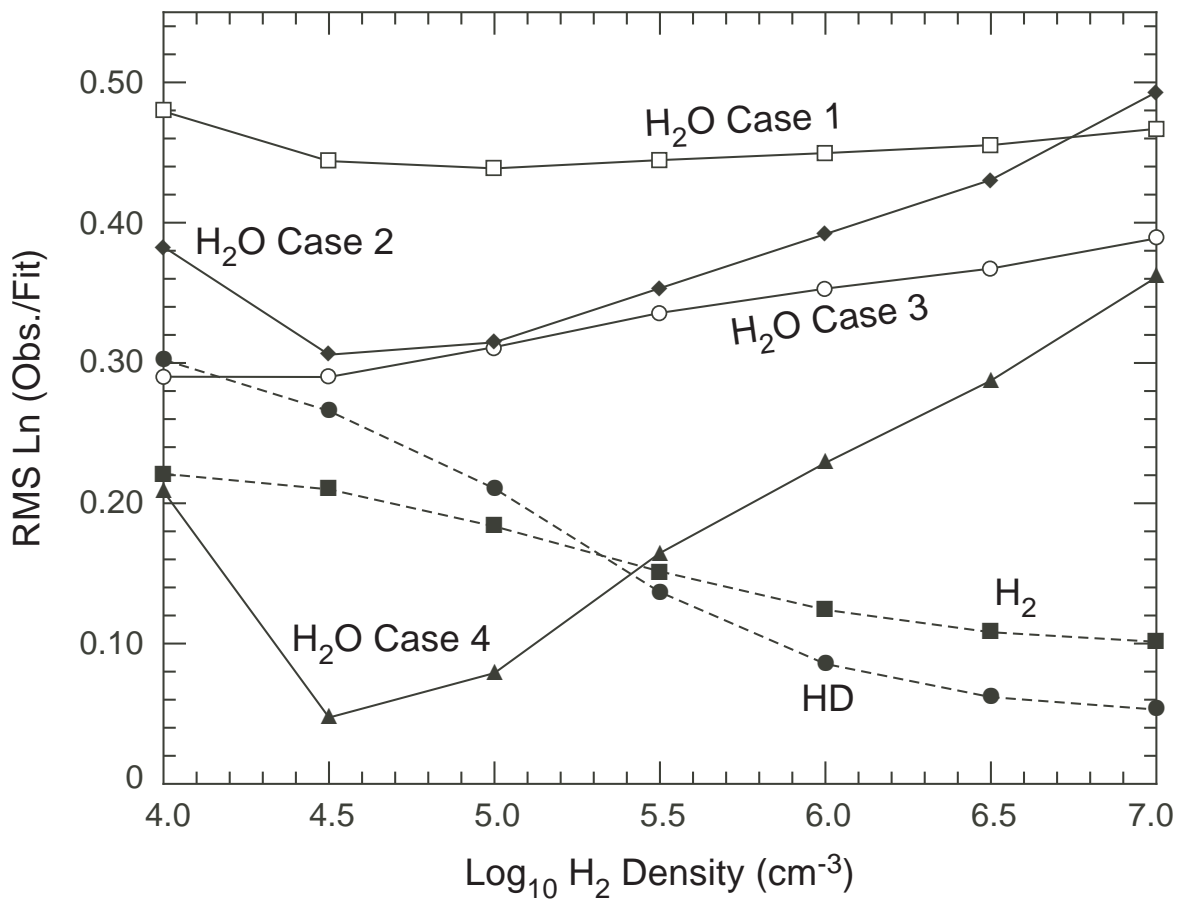}
\vspace{8mm}
\caption{Root mean square goodness-of-fit of the computed fluxes at each 
density for H$_2$ S(0)$\,-\,$S(7)
({\em dashed line-solid squares}); HD R(3) and R(4) ({\em dashed line-solid circles}); 
\water\ Case 1 -- the \water\ features listed in Table~2, with the exception of the 
34.55\um\ feature, and assuming an ortho-to-para
ratio of 3:1 ({\em open squares}); 
\water\ Case 2 -- the pure ortho- or para-\water\ lines at
29.84, 33.13, and 36.21\um\ and assuming an ortho-to-para
ratio of 3:1 ({\em solid diamonds});
\water\ Case 3 -- the \water\ features listed in Table~2, with the
exception of the 34.55\um\ feature, and assuming an ortho-to-para
ratio that has been allowed to vary to produce the best fit to the
data at each density ({\em open circle});
\water\ Case 4 -- the pure ortho- or para-\water\ lines at
29.84, 33.13, and 36.21\um\ and assuming an ortho-to-para
ratio that has been allowed to vary to produce the best fit to the
data at each density ({\em solid triangles}).}
\label{fig:Bestfitdensity}
\end{figure}

The good fit to the H$_2$ and HD data 
(see Fig.~\ref{fig:Bestfitdensity}) obtained with conditions characteristic of 
a $C$-type shock provide the best evidence for the coexistence of a non-dissociative
shock.  Based on the morphology of the Si$\,$II, Fe$\,$II, and H$_2$ emission,
it is tempting to suggest that the $C$-type shock emission arises in the entrained 
slower moving shocked material surrounding the high-velocity $J$-type shock 
gas which, itself, slows and transitions to a $C$-type shock near position P1.  

That the water emission follows closely the H$_2$ emission is in accord with
the predictions of $C$-type shock models.  Unfortunately, our inability to 
unambiguously determine the \mh\ density in the water-emitting region 
from the {\em Spitzer}/IRS data
prevents us from fully testing the other key prediction of the $C$-type shock models -- 
namely that the water abundance will be high 
(i.e., $\chi$(H$_2$O)$\,\simgt\,$10$^{-4}$) in the post-shocked gas.
Other observations provide general support for the density range preferred here. 
\citet{zhou91} have observed the NGC 2071 outflow in the CS
$J=\,$2-1, 5-4, 6-5, and 7-6 transitions.  Their best fit to the CS wing
emission yield densities
of 1$\,-\,$4 $\times$ 10$^5$\cmc.  Unfortunately, the varying
beam sizes with which these observations were conducted, ranging
between 11\asec\ and 24\asec, along with uncertainties in the gas 
temperature and the surface filling factors for the different transitions
contribute to the overall uncertainty in the density determination.
Nonetheless, if this range of densities applies to H$_2$O Peak 1 region,
then the best-fit water abundance would lie between approximately 
5$\times$10$^{-5}$ and 2$\times$10$^{-4}$ relative to H$_2$ which
still represents a significant abundance enhancement above that found
in quiescent molecular clouds \citep[e.g.,][]{snell00} and is consistent with
the predictions of $C$-type shock models.

\begin{figure}[b]
\centering
\includegraphics[scale=0.70]{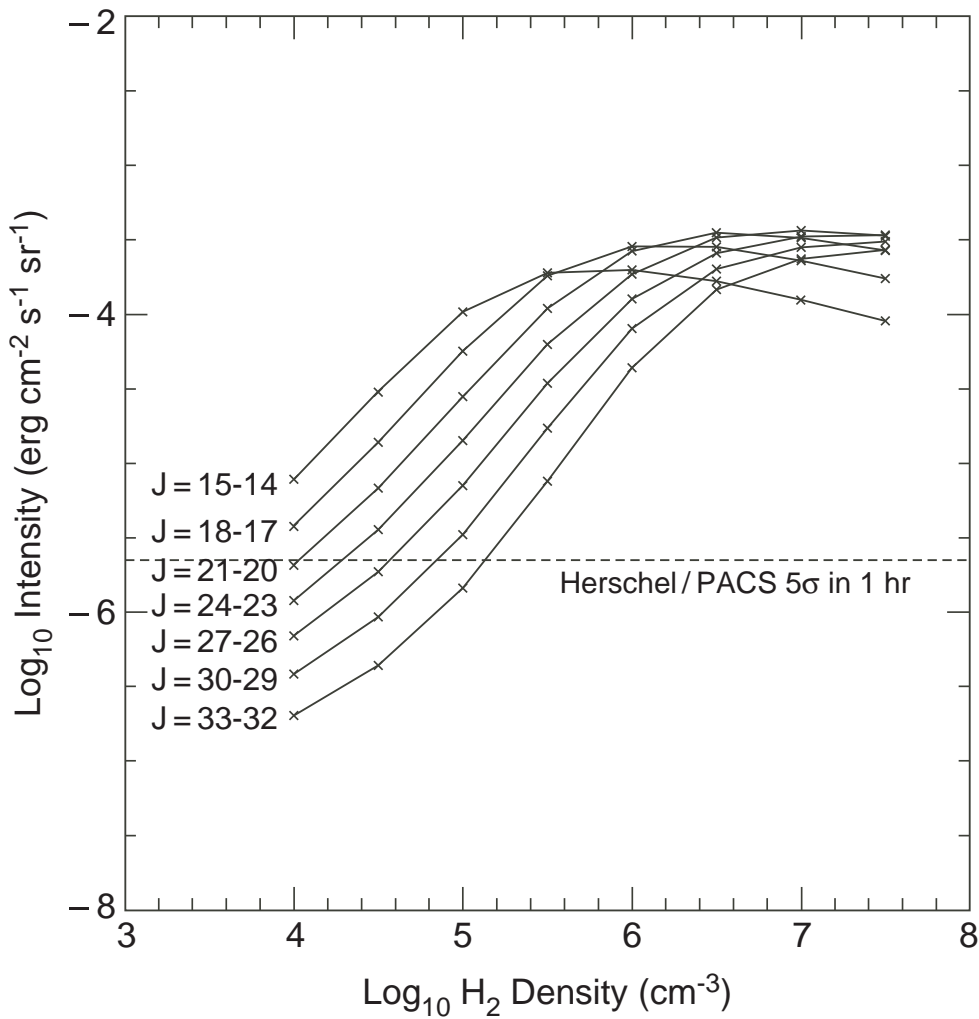}
\vspace{8mm}
\caption{Computed $^{12}$CO line intensities expected for a CO abundance of 10$^{-4}$ 
relative to H$_2$ (see text).  The dashed line corresponds to the {\em Herschel}/PACS 
spectral line sensitivity of $\sim\,$2$\times$10$^{-6}$ 
erg cm$^{-2}$ s$^{-1}$ sr $^{-1}$ (5$\sigma$ in 1 hour, corresponding to a flux of 
5$\times$10$^{-18}$ W m$^{-2}$ into a 9.7$\times$9.7\asec\ pixel).}  
\label{fig:CO-1}
\end{figure}

\begin{figure}[b]
\centering
\includegraphics[scale=0.70]{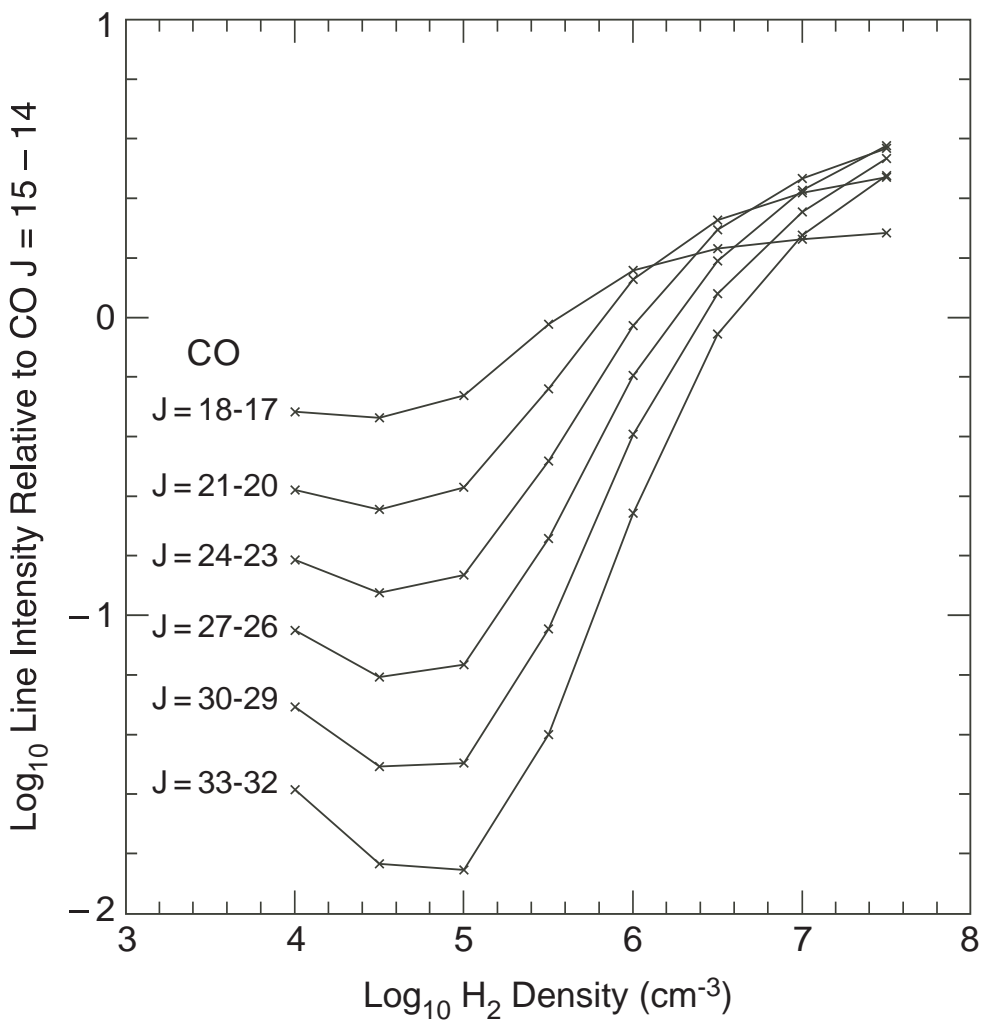}
\vspace{8mm}
\caption{Computed line intensity ratios of select $^{12}$CO high-$J$ rotational
transitions with respect to the $^{12}$CO $J=\,$15$-$14 line
measurable with the {\em Herschel Space Observatory}.  These line ratios are particularly
sensitive to \mh\ densities between 10$^5$ and about 3$\times$10$^6$~\cmc.}
\label{fig:CO-2}
\end{figure}

Future observations of far-IR high-$J$ CO lines -- planned for the Photodetector 
Array Camera and Spectrometer (PACS) instrument aboard the {\em Herschel 
Space Observatory} -- promise to yield a valuable means of estimating the 
\mh\ density in NGC 2071.   Given the gas temperatures and column densities 
derived from our {\em Spitzer} observations of the H$_2$ S(0)$-$S(7) rotational 
transitions,  we have computed the CO line intensities expected for a CO abundance 
of 10$^{-4}$ relative to H$_2$.  The sample results shown in Fig.~\ref{fig:CO-1} 
were obtained by means of a statistical equilibrium calculation in which we adopted 
the molecular data described in \citet{neufeld93}.  Fig.~\ref{fig:CO-1}
indicates that far-IR transitions of CO should be readily detectable with {\em Herschel} 
and that their strengths will provide a key probe of the gas density over exactly the 
desired range:  4$\,<\,$log$_{10}$(H$_2$/\cmc)$\,<\,$6.5.   Here, the dashed line 
corresponds to the {\em Herschel}/PACS spectral line sensitivity of $\sim\,$2$\times$10$^{-6}$ 
erg cm$^{-2}$ s$^{-1}$ sr $^{-1}$ (5$\sigma$ in 1 hour, corresponding to a flux of 
5$\times$10$^{-18}$ W m$^{-2}$ into a 9.7$\times$9.7\asec\ pixel).  
Because the CO lines are optically-thin, the predicted line intensities are proportional 
to the assumed CO abundance.  However, the far-IR line ratios (Fig.~\ref{fig:CO-2}) 
are independent of the assumed CO abundances and would provide a good estimate 
of densities in the range 5$\,<\,$log$_{10}$(H$_2$/\cmc)$\,<\,$6.5
without any required knowledge of the CO abundance.

The question of the distribution and abundance of the gas-phase water is important
not only as the answers test the validity of long-standing chemical predictions.
Many models of the larger ambient clouds take as a starting point certain molecular
enrichment due to the passage of  nearby shocks.  A recent analysis of 18 molecular
outflows by \citet{franklin08}
based on observations of the ground-state ortho-\water\ transition from 
{\em SWAS} and the $J =\,$1$-$0 transition of $^{12}$CO and $^{13}$CO obtained
with the Five College Radio Astronomy Observatory shows that the ortho-\water\
abundance (relative to H$_2$) in most outflows is between 10$^{-7}$ and 10$^{-6}$,
assuming the \water\ and CO emission arises in the same gas.  However, an
examination of the water abundance as a function of outflow velocity reveals 
a strong dependence; the water abundance rises with outflow velocity, reaching
abundances of $\sim\,$10$^{-4}$ at the highest velocities.  However, the mass
associated with the highest velocity emission is found to be small compared with
the total outflow mass, leading to the conclusion that only a small fraction of
the outflowing molecular gas has passed through shocks strong enough to fully
convert the gas-phase oxygen to water.  Using the measurements obtained here, 
combined with the data soon to be available from
{\em Herschel}, it will be possible to independently test this conclusion.

Finally, our potential detection of an \water\ ortho-to-para ratio that is not in
equilibrium at the gas temperature may hint at the origin of the water vapor emission
we detect.
The lowest energy level of para-H$_2$O is $\sim\,$34~K below that of 
ortho-H$_2$O and, as such, the ratio of the ortho and para populations 
in LTE is temperature dependent.   When water forms in the gas phase via exothermic 
reactions the energy released is much greater than this energy difference and the 
ortho-to-para ratio will reflect the 3:1 ratio of statistical weights between these
species.   When water forms on the surfaces of cold dust grains, it is believed 
that the energy generated by the chemical reaction is shared with the grain and the 
water molecules should equilibrate to an ortho-to-para ratio that reflects the 
grain temperature \citep[e.g.,][]{limbach06}.   If the grain temperature is 
below $\sim\,$50~K, then the ortho-to-para ratio will lie below 3:1.
In this regard, it has been known for many years that water in numerous 
cometary comae exhibits an ortho-to-para ratio below 3:1, which has been 
attributed to formation at temperatures of $\sim\,$25$-$30~K
\citep[e.g.,][]{mumma87, bonev07}.  

Our measurements, while uncertain, potentially limit the ortho-to-para ratio
to between 1.5--2.5, or equivalent temperatures of $\sim\,$20$-$30~K.  If the water
is produced entirely in shocks via the well known sequence of neutral-neutral reactions, 
the ortho-to-para ratio should be 3:1.   However, if the water enters the gas phase via 
sputtering from cold grain surfaces, then the ratio would 
reflect equilibrium at the grain temperature.   A mixture of both sputtering and 
gas-phase formation cannot be ruled out and thus this measurement represents an 
upper limit to the ortho-to-para ratio of water ice.  If our measurement is 
confirmed, this would therefore suggest that a significant amount of the \water\
emission we detect has its source in water vapor produced
from grain sputtering and not via gas-phase reactions.

\vspace{14mm}

\noindent{\em Acknowledgements}:~~G.J.M. gratefully acknowledges the financial 
support of NASA grant\\ NNG06GB30G from the Long Term Space Astrophysics 
(LTSA) Research Program and through an award issued by JPL/Caltech
(Support Agreement no.~1265773).
This work is based on observations made with the {\em Spitzer Space Telescope}, 
which is operated by JPL/Caltech under NASA contract 1407.

\clearpage

\renewcommand{\baselinestretch}{1.00}

\renewcommand{\baselinestretch}{1.02}

\pagebreak

\phantom{0}
\vspace{9mm}

\normalsize

\begin{center}
Table~1.~Peak Intensity for Each Mapped Transition \\*[1.4mm]
\begin{tabular}{lccc} \hline
\rule{0mm}{6mm} &  & Rest & Peak \\*[-1.0mm]
~Species~~~ & ~~~~Transition~~~~ & ~~Wavelength~~ & ~~~Intensity~~~ \\*[-0.5mm]
 &  & \um\ & 10$^{-6}$ erg cm$^{-2}$ s$^{-1}$ sr$^{-1}$ \\*[0.6mm] \hline
\rule{0mm}{6mm} ~~~H$_2$O & 7$_{25}$ -- 6$_{16}$ & 29.837 & \phantom{2}4.39 \\*[2mm]
                      & 6$_{34}$ -- 5$_{05}$ & 30.899 & \phantom{2}7.08 \\
                      & 8$_{54}$ -- 7$_{43}$ & 30.871 &  \\*[2mm]
                      & 7$_{34}$ -- 6$_{25}$ & 34.549 & \phantom{2}7.08 \\*[4mm]
~~~H$_2$ & S(0) & 28.221 & 21.10 \\*[0.3mm]
      & S(1) & 17.035 & 189.31\phantom{0} \\*[0.3mm]
      & S(2) & 12.279 & 1721\phantom{0000} \\*[0.3mm]
      & S(3) & \phantom{1}9.665 & 1421\phantom{0000} \\*[0.3mm]
      & S(4) & \phantom{1}8.025 & 1200\phantom{0000} \\*[0.3mm]
      & S(5) & \phantom{1}6.909 & 3300\phantom{0000} \\*[0.3mm]
      & S(6) & \phantom{1}6.109 & 1746\phantom{0000} \\*[0.3mm]
      & S(7) & \phantom{1}5.511 & 1800\phantom{0000} \\*[4mm]
~~~HD    & R(3) & 28.502 & \phantom{2}3.37 \\*[0.3mm]
      & R(4) & 23.034 & \phantom{2}6.20 \\*[4mm]
~~~S$\,$I    & $^3$P$_1$ -- $^3$P$_2$ & 25.249 & 201.70\phantom{0} \\*[4mm]
~~~Fe$\,$II & $a^6$D$_{7/2}$ -- $a^6$D$_{9/2}$ & 25.988 & 40.84 \\*[4mm]
~~~Si$\,$II & $^2$P$_{3/2}$ -- $^2$P$_{1/2}$ & 34.815 & 55.41 \\*[0.8mm] \hline 
\end{tabular}
\end{center}

\eject

\pagebreak

\phantom{0}
\vspace{-1mm}

\normalsize

\begin{center}
Table~2.~~Beam-Averaged Line Fluxes for the H$_2$O Features Toward 
H$_2$O Peak 1\\*[1.4mm]
\begin{tabular}{ccccc} \hline

\rule{0mm}{6mm} & & &  & Extinction-Corrected \\
H$_2$O & ~~Ortho (o)/~~ &  & Rest & ~~Line Flux $\times$ 10$^{-21}$~~ \\
~~~Feature~~~ & Para (p) & ~~~~Transition~~~~& 
 ~~~~Wavelength~~~~ & into 15\asec\ Beam$^{\,\rm a}$ \\
\um\ &  &  & \um\ & W cm$^{-2}$ \\ \hline
\rule{0mm}{6mm}36.21 & p & 6$_{24}$ -- 5$_{15}$ &  36.212 & 1.57 \\*[3mm]
35.92 & o & 6$_{52}$ -- 5$_{41}$ & 35.938 & 0.70 \\*[0.5mm]
                  & p & 6$_{51}$ -- 5$_{42}$ & 35.904 &  \\*[3mm]
35.67 & o & 8$_{45}$ -- 7$_{34}$ & 35.669 & 0.55 \\*[3mm]
35.45 & p & 5$_{33}$ -- 4$_{04}$ & 35.472 & 2.43 \\*[0.5mm]
     & o & 7$_{43}$ -- 6$_{34}$ & 35.428 &  \\*[3mm]
34.55 & o & 7$_{34}$ -- 6$_{25}$ & 34.549 & 2.02 \\*[3mm]
33.13 & p & 7$_{53}$ -- 6$_{42}$ & 33.127 & 0.64 \\*[3mm]
32.99 & o & 16$_{1\,16}$ -- 15$_{0\,15}$ & 33.042 & 1.42 \\*[0.5mm]
   & p  & 16$_{0\,16}$ -- 15$_{1\,15}$ & 33.042 &  \\*[0.5mm]
   & o  & 6$_{61}$ -- 5$_{50}$ & 33.005 & \\*[0.5mm]
   & p & 6$_{60}$ -- 5$_{51}$ & 33.005 & \\*[0.5mm]
   & o  & 7$_{52}$ -- 6$_{43}$ & 32.991 & \\*[0.5mm]
   & o  & 15$_{1\,14}$ -- 14$_{2\,13}$ & 32.965 &  \\*[0.5mm]
   & p  & 15$_{2\,14}$ -- 14$_{1\,13}$ & 32.960 &  \\*[3mm]
31.75 & o & 4$_{41}$ -- 3$_{12}$ & 31.772 & 0.70 \\*[0.5mm]
          & p & 8$_{44}$ -- 7$_{35}$ & 31.738 &   \\*[3mm]
30.89 & o & 6$_{34}$ -- 5$_{05}$ & 30.899 & 1.82 \\*[0.5mm]
      & o & 8$_{54}$ -- 7$_{43}$ & 30.871 &  \\*[3mm]
30.53 & p & 7$_{62}$ -- 6$_{51}$ & 30.529 & 0.63 \\*[0.5mm]
      & o & 7$_{61}$ -- 6$_{52}$ & 30.526 & \\*[3mm]
29.84 & o & 7$_{25}$ -- 6$_{16}$ & 29.837 & 1.27 \\*[0.8mm] \hline 
\end{tabular}
\end{center}

\vspace{-4.0mm}

\begin{list}{}{\leftmargin 0.73in \rightmargin 0.73in \itemindent -0.13in}
\item {$^{\rm a}$}~At the spectral resolving power of {\em Spitzer}/IRS, some
H$_2$O features may be blends of two or more H$_2$O transitions.  The line
fluxes listed here represent the {\em total} flux for each H$_2$O feature.\\*[-5mm]
\end{list}

\eject

\pagebreak

\phantom{0}
\vspace{-1mm}

\normalsize

\begin{center}
Table~3.~~Beam-Averaged Line Fluxes for Key Species Toward H$_2$O Peak 1\\*[1.4mm]
\begin{tabular}{lccc} \hline
\rule{0mm}{6mm} &  &  & Extinction-Corrected \\
  &  & Rest & ~~~Line Flux $\times$ 10$^{-21}$~~~ \\
~~Species~~ & ~~~~~Transition~~~~~ & ~~~Wavelength~~~ &  into 15\asec\ Beam \\
  & & \um\ & W cm$^{-2}$ \\*[1mm] \hline
\rule{0mm}{6mm}~~~~~H$_2$~~ & S(7) & ~~5.5112 &  275.2 \\*[1.4mm]
~~~~~H$_2$ & S(6) & ~~6.1086 &  110.9 \\*[1.4mm]
~~~~~H$_2$ & S(5) & ~~6.9095 &  513.0 \\*[1.4mm]
~~~~~H$_2$ & S(4) & ~~8.0251 &   207.0 \\*[1.4mm]
~~~~~H$_2$ & S(3) & ~~9.6649 &  474.4 \\*[1.4mm]
~~~~~H$_2$ & S(2) & 12.2786 &  138.1 \\*[1.4mm]
~~~~~H$_2$ & S(1) & 17.0348 &  ~~68.3 \\*[1.4mm]
~~~~~HD & R(4) & 23.0338 &  ~~~~2.2 \\*[1.4mm]
~~~~~OH & ~~~~$^2\Pi_{3/2}$ 23/2$^+$ - 21/2$^-$~~~~ & 25.0351 & ~~~~~~0.41 \\*[1.4mm]
~~~~~OH & $^2\Pi_{3/2}$ 23/2$^-$ - 21/2$^+$ & 25.0899 & ~~~~~~0.48 \\*[1.4mm]
~~~~~[S$\,$I] & $^3$P$_1$ -- $^3$P$_2$ & 25.2490 &  ~~78.8 \\*[1.4mm]
~~~~~[Fe$\,$II] & $a^6$D$_{7/2}$ -- $a^6$D$_{9/2}$ & 25.9883 &  ~~~~7.6\\*[1.4mm]
~~~~~OH & $^2\Pi_{3/2}$ 21/2$^-$ - 19/2$^+$ &  27.3935 & ~~~~~~0.13 \\*[1.4mm]
~~~~~OH & $^2\Pi_{3/2}$ 21/2$^+$ - 19/2$^-$ &  27.4546 & ~~~~~~0.58 \\*[1.4mm]
~~~~~OH & $^2\Pi_{1/2}$ 19/2$^-$ - 17/2$^+$ &  27.6516 & ~~~~~~0.31\\*[1.4mm]
~~~~~OH & $^2\Pi_{1/2}$ 19/2$^+$ - 17/2$^-$ &  27.6967 & ~~~~~~0.26 \\*[1.4mm]
~~~~~H$_2$ & S(0) & 28.2188 &  ~~~~9.1 \\*[1.4mm]
~~~~~HD & R(3) & 28.5020 &  ~~~~2.3 \\*[1.4mm]
~~~~~OH & $^2\Pi_{3/2}$ 19/2$^+$-17/2$^-$  & 30.2772 & ~~~~~~0.25 \\*[1.4mm]
~~~~~OH & $^2\Pi_{3/2}$ 19/2$^-$-17/2$^+$  & 30.3459 & ~~~~~~0.60 \\*[1.4mm]
~~~~~OH & $^2\Pi_{1/2}$ 17/2$^+$-15/2$^-$ & 30.6573  & ~~~~~~0.33 \\*[1.4mm]
~~~~~OH & $^2\Pi_{1/2}$ 17/2$^-$-15/2$^+$ & 30.7063  & ~~~~~~0.37 \\*[1.4mm]
~~~~~[S$\,$III] & $^3$P$_1$ -- $^3$P$_0$  & 33.4800 &  ~~~~1.5\\*[1.4mm]
~~~~~[Si$\,$II] & $^2$P$_{3/2}$ -- $^2$P$_{1/2}$ & 34.8152 & ~~18.7 \\*[1.4mm] \hline
\end{tabular}
\end{center}

\eject

\pagebreak

\phantom{0}
\vspace{9mm}

\normalsize

\begin{center}
~~~~~~Table~4.~Best-Fit Physical Conditions Within H$_2$O Peak 1 Inferred 
from H$_2$ S(0)-S(1) Lines \\*[1.4mm]
\begin{tabular}{lccccccc} \hline
\rule{0mm}{5.8mm} & \multicolumn{7}{c}{Log$_{10}$ H$_2$ Density (cm$^{-3}$)} \\*[1mm] \cline{2-8}
\rule{0mm}{6mm}~~~~Parameter & 4.0 & 4.5 & 5.0 & 5.5 & 6.0 & 
    6.5 & 7.0 \\*[1.1mm] \hline
\rule{0mm}{6.4mm}Warm Component:\hspace{1.3in} & & & & & & & \\*[2mm]
~~~Temperature (K)\dotfill & ~~377 & ~~361 & ~~352 & ~~344 & 
     ~~336 & ~~330 & ~~328 \\*[2.4mm]
~~~Log$_{10}$ N(H$_2$) (cm$^{-2}$) -- measured\dotfill & 20.60 & 20.61 & 
     20.61 & 20.60 & 20.59 & 20.59 & 20.58 \\*[2.4mm]
~~~Log$_{10}$ N(H$_2$) (cm$^{-2}$) -- theory\dotfill & 20.54 & 20.81 & 
     21.06 & 21.32 & 21.57 & 21.83 & 22.08 \\*[2.4mm]
~~~H$_2$ Ortho-to-Para Ratio\dotfill & ~0.69 & ~0.71 & ~0.68 & ~0.63 & 
     ~0.58 & ~0.55 & ~0.53 \\*[2.4mm]
~~~Beam Filling Factor\dotfill & 1.0~ & ~0.64 & ~0.35 & ~0.19 & ~0.10 &
      ~0.06 &~ 0.03 \\*[2.4mm]
~~~Mass of Warm H$_2$ ($\times$ 0.1 M$_{\odot}$)\dotfill & ~0.04 & 
      ~0.07 & ~0.12 & ~0.21 & ~0.38 & ~0.68 & ~1.22 \\*[3mm]
\rule{0mm}{6mm}Hot Component: & & & & & & & \\*[2.4mm]
~~~Temperature (K)\dotfill & 2111 & 1621 & 1341 & 1169 & 1070 & 1024 & 1006 \\*[2.4mm]
~~~Log$_{10}$ N(H$_2$) (cm$^{-2}$) -- measured\dotfill & 19.31 & 19.38 & 
    19.47 & 19.58 & 19.66 & 19.71 & 19.73 \\*[2.4mm]
~~~Log$_{10}$ N(H$_2$) (cm$^{-2}$) -- theory\dotfill & 20.13 & 20.44 & 
    20.74 & 21.02 & 21.29 & 21.55 & 21.81 \\*[2.4mm]
~~~H$_2$ Ortho-to-Para Ratio$^{\rm \,a}$\dotfill & ~3.45 & ~3.58 & ~3.68 & ~3.68 & 
    ~3.62 & ~3.56 & ~3.53 \\*[2.4mm]
~~~Beam Filling Factor\dotfill & ~0.15 & ~0.09 & ~0.05 & ~0.04 & 
      ~0.02 &  ~0.01 &  ~~0.008 \\*[2.4mm]
~~~Mass of Hot H$_2$ ($\times$ 0.1 M$_{\odot}$)\dotfill & ~0.01 & 
      ~0.03 & ~0.06 & ~0.11 & ~0.20 & ~0.36 & ~0.65  \\*[1.8mm] \hline
\end{tabular}
\end{center}

\vspace{-6.5mm}

\begin{list}{}{\leftmargin 0.20in \rightmargin 0.00in \itemindent -0.13in}
\item $^{\rm a}\;$Errors on the H$_2$ ortho-to-para ratio are hard to estimate,
but the best-fit values for the hot component are probably consistent with the
LTE value of 3.
\end{list}

\eject

\clearpage

\phantom{0}

\clearpage

\clearpage

\begin{figure}[t]
\centering
\includegraphics[scale=0.80]{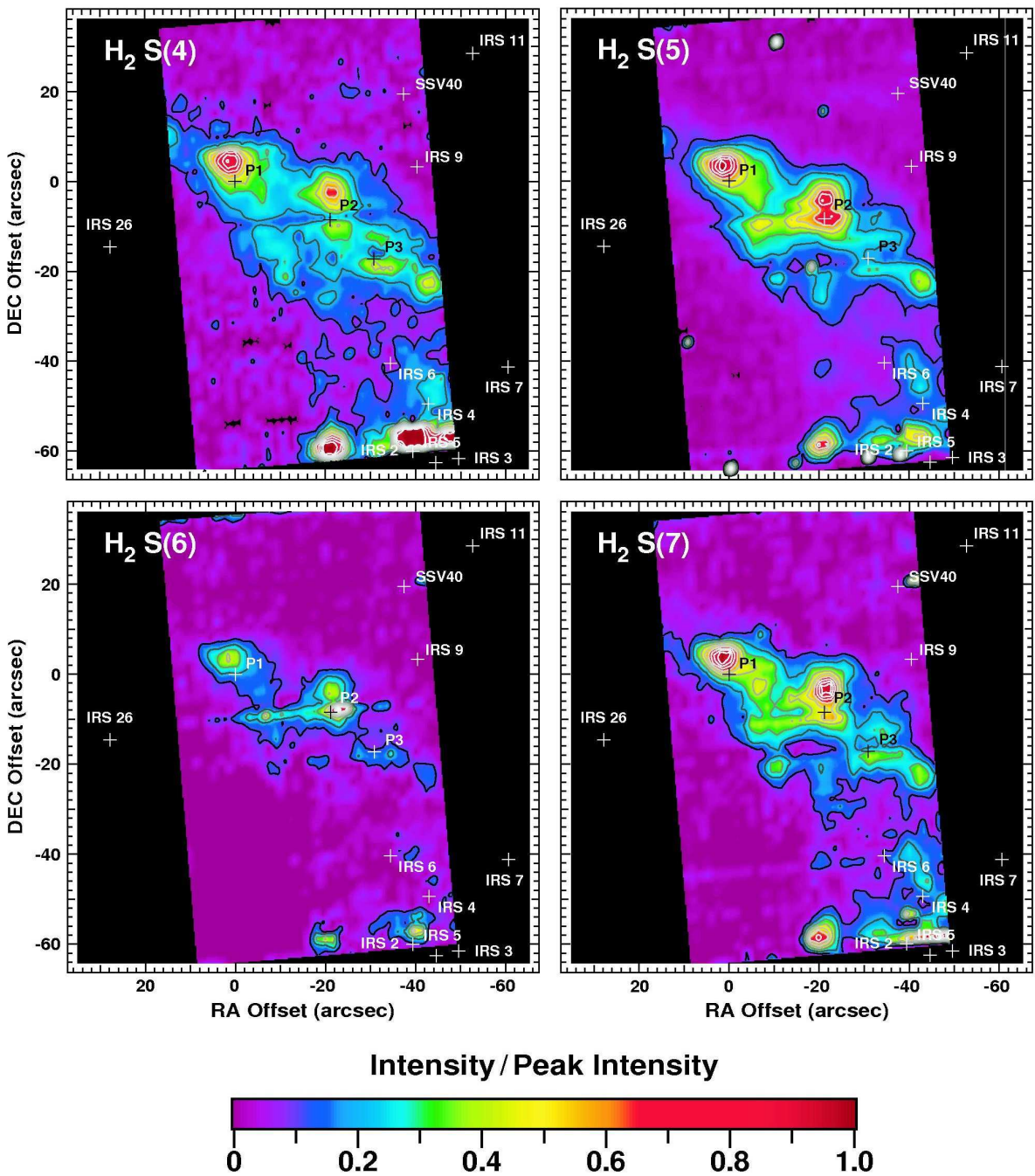}
\vspace{3mm}
\caption{Same as Fig.~2, except for H$_2$ S(4) through S(7) emission.}
\label{fig:map2}
\end{figure}

\clearpage

\clearpage

\begin{figure}[t]
\vspace{1.2in}
\centering
\includegraphics[scale=0.80]{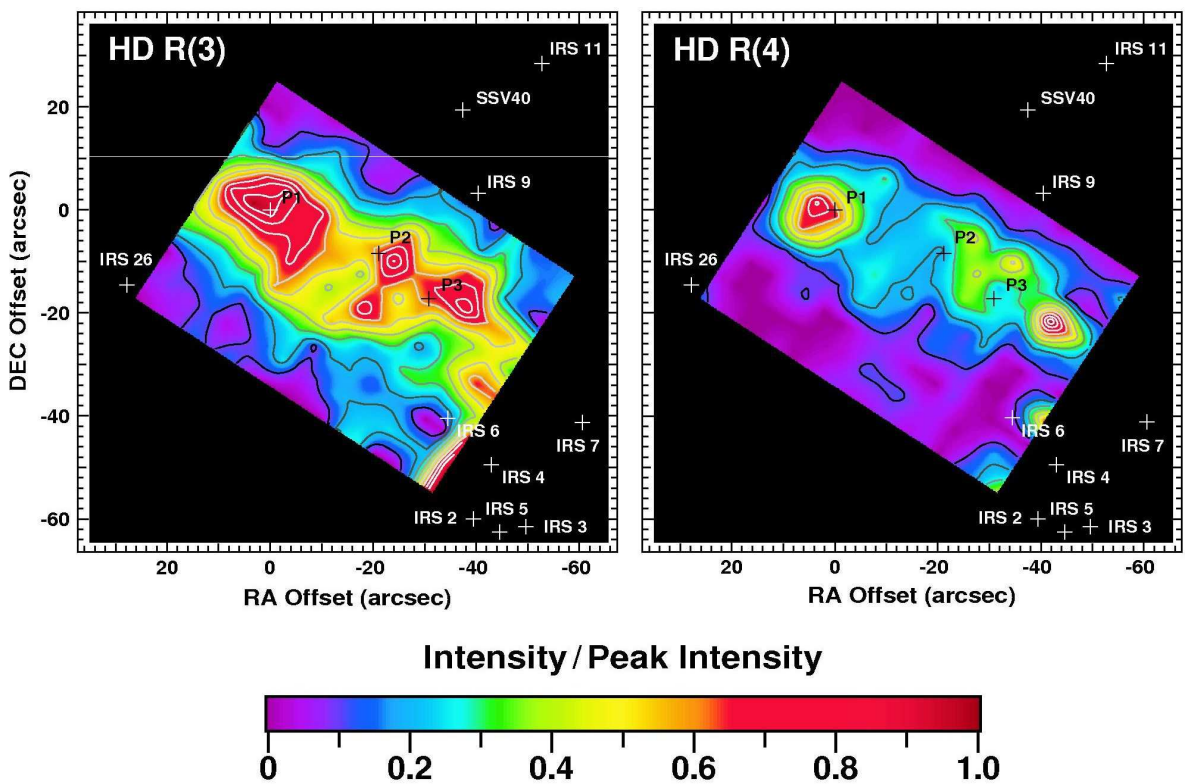}
\vspace{7mm}
\caption{Same as Fig.~2, except for HD R(3) 28.502\um\ and R(4) 23.034\um\ 
pure rotational
emission.}
\label{fig:map5}
\end{figure}

\clearpage

\clearpage

\clearpage

\clearpage

\clearpage

\clearpage

\clearpage

\clearpage

\clearpage

\clearpage

\end{document}